\documentclass[a4paper,11pt]{article}
\pdfoutput=1 

\usepackage{jheppub} 

\usepackage[T1]{fontenc} 
\usepackage{soul}
\usepackage{color}

\usepackage{empheq}
\usepackage{mathrsfs} 
\usepackage{url}
\usepackage[utf8]{inputenc}
\usepackage[toc,page]{appendix}
\usepackage[autostyle]{csquotes}
\usepackage{slashed}
\usepackage{amssymb}
\usepackage{graphicx}
\usepackage{dirtytalk}
\usepackage{longtable}
\usepackage{array}
\usepackage{bm} 
\usepackage{amsmath}
\usepackage{nccmath}
\usepackage{cancel}
\usepackage{mathtools}
\usepackage{amsfonts}
\usepackage{amssymb}
\usepackage[labelfont=bf]{caption}
\usepackage{paralist} 
\usepackage{hyperref}
\usepackage{cleveref}

\title{\boldmath Seeley-DeWitt coefficients in $\mathcal{N}=2$ Einstein-Maxwell supergravity theory and  logarithmic corrections to $\mathcal{N}=2$ extremal black hole entropy}

\author{Sudip Karan,}
\author{Gourav Banerjee}
\author{and Binata Panda}
\affiliation{Department of Physics, Indian Institute of Technology (Indian School of Mines),\\ Dhanbad, Jharkhand-826004, India}

\emailAdd{sudip.karaan@gmail.com}
\emailAdd{gourav9124@gmail.com}
\emailAdd{binata@iitism.ac.in}

\abstract{We investigate the heat kernel method for one-loop effective action following the  Seeley-DeWitt expansion technique of heat kernel with Seeley-DeWitt coefficients. We also review a general approach of computing the Seeley-DeWitt coefficients in terms of background or geometric invariants. We, then consider the Einstein-Maxwell theory embedded in minimal $\mathcal{N}=2$  supergravity in four dimensions and compute the first three Seeley-DeWitt coefficients of the kinetic operator of the bosonic and the fermionic fields in an arbitrary background field configuration. We find the applications of these results in the computation of logarithmic corrections to Bekenstein-Hawking entropy of the extremal Kerr-Newman, Kerr and Reissner-Nordstr\"om black holes in minimal $\mathcal{N}=2$ Einstein-Maxwell supergravity theory following the quantum entropy function formalism.}

\begin{document} 
\maketitle
\flushbottom

\section{Introduction}\label{intro}

Bekenstein-Hawking entropy formula is an approximated formula in two-derivative classical general relativity where one ignores the higher derivative terms in the action. Remarkably, this formula is very simple and universal in the sense that it is completely governed by horizon area, does not bother about the past history or detailed matter content of the particular black hole. In string theory, the Bekenstein-Hawking formula is subjected to two types of corrections — higher curvature corrections and quantum corrections \cite{Mandal:2010cj}. The higher curvature corrections arise due to considering higher derivative terms in the classical action, are captured by the Wald's generalization of Bekenstein-Hawking entropy.
On the other hand, the quantum corrections are important for black holes of a size comparable to Planck size. The Bekenstein-Hawking  formula works perfectly for macroscopically large black holes ($A_{H}\gg1$)\footnote{Actually, large black holes have horizon area $A_{H}\gg l_P^2$ ($l_P$ is the Planck length). But in this paper we are considering the units $\hbar = c = k_{B} = 1$.} and is expected to receive quantum corrections as one lowers the size of the black hole. In this work, we are interested in a certain class of quantum corrections as discussed below.  

Recent years there have been remarkable research activities\footnote{See \cite{Setare:2003pj} for a list of references.} in order to find quantum corrections to black hole entropy formula and the quantum-corrected entropy received the following form,
\begin{equation}
	S_{\text{BH}} = \frac{A_{H}}{4G_N}+ c\thinspace \text{ln}\thinspace \frac{A_{H}}{G_N} + \text{constant} + \mathcal{O}(A_{H}^{-1}).
\end{equation}
Here $c\thinspace \text{ln}\thinspace \frac{A_{H}}{G_N}$ is the \textit{logarithmic correction}, proportional to the logarithmic of the size of the black hole, with a proportionality constant $c$ and $\mathcal{O}(A^{-1})$ are power-law corrections to the leading order term in the black hole entropy formula. Although the total quantum corrections will depend on the details of the UV completion of the low energy theory, the logarithmic corrections can be determined from the knowledge of only low energy modes. These corrections can be computed from one-loop effective action of the massless fields in the particular theory. These logarithmic corrections are universal in this sense that they appear in any generic theory of gravity. But $c$ does not have the same sort of universality: it can be determined only from low energy modes, i.e., the spectrum of massless fields present in the theory, and are unconcerned about UV completion of the theory \cite{Banerjee:2011oo,Banerjee:2011pp,Sen:2012qq,Sen:2012rr,Bhattacharyya:2012ss,Chowdhury:2014np,Gupta:2014ns,Keeler:2014nn,Charles:2015nn,Larsen:2015nx,Castro:2018tg}. These correction terms are important as they provide us with the non-trivial information about the microstates of the black holes and hence providing a valuable testing ground for any candidate theory of quantum gravity. We still consider macroscopically large black holes to make the logarithmic terms as leading order corrections \cite{Banerjee:2011oo,Banerjee:2011pp}.

Quantum entropy function formalism \cite{Sen:2008wa,Sen:2009wb,Sen:2009wc} is used in \cite{Banerjee:2011oo,Banerjee:2011pp,Sen:2012rr,Sen:2012qq,Chowdhury:2014np,Gupta:2014ns} to compute the logarithmic corrections to the entropy of supersymmetric extremal black holes. This procedure uses the presence of $\text{AdS}_2$ factors in the near horizon geometry of extremal black holes and rules of $\text{AdS}_2/\text{CFT}_1$ correspondence. However, it does not depend on supersymmetry and can be applied to any extremal black holes \cite{Bhattacharyya:2012ss}. Computation of the logarithmic corrections involves finding the one-loop effective action of the massless fields in the near horizon geometry of black hole. Evaluation of effective action was investigated by many different methods, but we consider the most potent and traditional heat kernel method (discussed in \cref{ghkm}). The main advantage of this method is that it is explicitly covariant, as well as the most suitable tool to regularize the UV divergence of the one-loop effective action. Further, in order to compute the heat kernel for one-loop effective action we  pursue the Seeley-DeWitt expansion technique \cite{Seeley:1966tt,Seeley:1969uu,DeWitt:1965ff,DeWitt:1967gg,DeWitt:1967hh,
	DeWitt:1967ii,Duff:1977vv,Christensen:1979ww,Christensen:1980xx,Duff:1980yy,Birrel:1982zz,Gilkey:1984xy,Duff:2011yz}, followed by a general approach to compute the Seeley-DeWitt coefficients for fluctuations of gauge, graviton and matter fields around an arbitrary background field configuration demonstrated in \cite{Vassilevich:2003ll}. In this approach,
the Seeley-DeWitt coefficients are computed for an arbitrary
field configuration without imposing any limitations on the background geometry. Moreover, they are expressed in terms of local invariants calculated from background fields and their covariant derivatives. So one can employ these results for any arbitrary background field configuration.

In comparison, an incomplete list of many other approaches to compute heat kernel includes some perturbative expansion methods,\footnote{ E.g., see section 8 of \cite{Vassilevich:2003ll}} eigenfunction expansion method \cite{Banerjee:2011oo,Banerjee:2011pp,Sen:2012qq}, a group theoretic approach \cite{David:2010xn,Gopakumar:2011xi,Lovrekovic:2016ni}, quasinormal mode approach \cite{Denef:2010nc}, etc. But most of these approaches are either  background geometry dependent or model dependent or associated with some complex computations. Of these the  methods followed in \cite{Banerjee:2011oo,Banerjee:2011pp,Sen:2012qq} and \cite{Charles:2015nn,Larsen:2015nx,Castro:2018tg} are closest to the one advocated in this work; so  below we have given a detailed comparison between the methods. The advantages of the current approach are also outlined below \cref{115}.   
The eigenfunction  method to calculate the Seeley-DeWitt coefficients followed in \cite{Banerjee:2011oo,Banerjee:2011pp,Sen:2012qq} involves explicit knowledge of all the eigenfunctions and eigenvalues of the kinetic operator operating on the bosonic and the fermionic fields of the corresponding matter multiplet in the $\text{AdS}_2\times S^2$ background geometry. This is quite complicated and restricted to this type of geometry only. In the absence of rotational symmetry of the background geometry, the computation of eigenfunctions of the kinetic operator is  indeed a difficult task. Whereas the Seeley-DeWitt expansion approach involves expansion of the heat kernel in powers of small proper time and is applicable for any arbitrary background geometry beyond $\text{AdS}_2\times S^2$. Here
the Seeley-DeWitt coefficients are expressed in terms of geometric invariants and can be applied for any arbitrary background. 
Again, the technique
followed by Larsen et al. in \cite{Charles:2015nn,Larsen:2015nx,Castro:2018tg} is kind of hybrid — it is an extension of the general Seeley-DeWitt approach, where computation has been done by redefining fields present in the theory which induces loss of generality within the approach. 

The technical aim of this paper is twofold — first, following the general Seeley-DeWitt approach \cite{Vassilevich:2003ll} (reviewed in \cref{HKE}) we compute necessary Seeley-DeWitt coefficients of the quadratic order field fluctuations in the minimal  $\mathcal{N}=2$ Einstein-Maxwell supergravity theory (EMSGT) and then find their useful applications in the logarithmic correction to the Bekenstein-Hawking entropy formula for extremal black holes. The results for the fermionic sector of the minimal $\mathcal{N}=2$ EMSGT are newly computed for an arbitrary background, while the bosonic sector results are reviewed and recalculated from \cite{Bhattacharyya:2012ss}. These Seeley-DeWitt coefficient results for the minimal $\mathcal{N}=2$ EMSGT in four dimensions agree with eigenfunction approach results of \cite{Sen:2012qq} for $\text{AdS}_2\times S^2$ background, as well as with the hybrid approach results of \cite{Charles:2015nn}.
Then the logarithmic corrections to the entropy of extremal black holes in  minimal $\mathcal{N}=2$ EMSGT are computed using the quantum entropy function formalism. The result is stated in \cref{68}. The result is new in the literature for extremal Kerr- Newman black holes. Any microscopic theory explaining the entropy of these black holes should reproduce the result specified in \cref{68}. With appropriate limits \cref{68} produces the Kerr and Reissner-Nordstr\"om results which are consistent with \cite{Sen:2012rr} and \cite{Sen:2012qq}, respectively.

The rest of this paper is structured as follows. In \cref{heat kernel}, we will discuss the heat kernel method in four-dimensional Euclidean gravity. In this context, we will express the one-loop effective action of our theory in terms of heat kernels of the corresponding kinetic operator, and also introduce Seeley-DeWitt coefficients in the short-time expansion of these heat kernels. In \cref{HKE}, we will review the general approach, necessary formulae and steps for calculating the Seeley-DeWitt coefficients. In \cref{EMSGT}, we consider Einstein-Maxwell theory embedded in minimally coupled $\mathcal{N}=2$ supergravity in 4D and briefly explore the basic structure, equations of motion and important identities of the resulting minimally coupled $\mathcal{N}=2, d=4$ EMSGT. In \cref{HKIEMSGT}, we divide the total field contents of the minimally coupled $\mathcal{N}=2, d=4$ EMSGT into bosonic and fermionic sectors, then employ the general approach to calculate the first three Seeley-DeWitt coefficients for the two sectors separately.\footnote{We actually compute results for the fermionic sector, while we review \cite{Bhattacharyya:2012ss} for the bosonic sector.} In \cref{app}, we will use the results for Seeley-DeWitt coefficients to compute logarithmic corrections to the entropy of extremal Kerr-Newman, Reissner-Nordstr\"om and Kerr black holes in minimally coupled $\mathcal{N}=2$  EMSGT following quantum entropy function formalism, and comment on the consistency of the results.
\section{Heat kernel in 4D Euclidean gravity}\label{heat kernel}
In this section, we briefly summarize the heat kernel approach in the context of 4D Euclidean gravity theory. Effective action plays a crucial theoretical role in quantum field theory and quantum gravity since it encodes almost all information about a quantum system. Unfortunately, it is associated with some mathematical pathologies and hence very difficult to compute. Heat kernel method is found to be a useful tool in this purpose.
\subsection{Euclidean path integrals: the set-up}
We start with the partition function $Z$ 
in Euclidean path integral representation \cite{Gibbons:1977ta},
\begin{align}
	& Z= \int D[g,\xi]\text{exp}\left(-\mathcal{S}[g,\xi] \right),\label{200}\\
	& \mathcal{S}[g,\xi] = \int_{\mathcal{M}} d^4x\sqrt{\text{det}\thinspace g} \thinspace\mathcal{L}[g,\xi].\label{201}
\end{align}
$\mathcal{S}[{g},\xi]$ is the Euclidean action of the matter fields $\xi$ propagating through a geometry described by the metric ${g}$ over an arbitrary manifold $\mathcal{M}$; $D[{g},\xi]$ suggests that we have to carry out the functional  integration  \eqref{200} over all possible matter fields $\xi$ and metric $g$. 

The leading contribution to the path integral \eqref{200} is expected to arise from the metric and the matter fields that are near the background fields $\bar{g}$ and $\bar{\xi}$. Here $(\bar{g}, \bar{\xi})$ extremize our action \eqref{201}, i.e.\ they are the solution of the classical field equations with proper boundary conditions and periodicity. At this stage, one can express ${g}$ and $\xi$ in terms of their fluctuations around the background as
\begin{equation}\label{202}
	{g} = \bar{g}+ \tilde{{g}},\enspace \xi = \bar\xi + \tilde{\xi},
\end{equation}
and expand the action in Taylor series around the classical solution $(\bar{g}, \bar{\xi})$
,
\begin{equation}\label{203}
	\mathcal{S}[{g},\xi] = \mathcal{S}[\bar{g}, \bar{\xi}]+ \mathcal{S}_2[\tilde{{g}},\tilde{\xi}]+ \text{higher order terms},
\end{equation}
where $\mathcal{S}[\bar{g}, \bar{\xi}]$ is the action of the classical background fields $\bar{g}$ and $\bar{\xi}$; $\mathcal{S}_2[\tilde{{g}},\tilde{\xi}]$ is the quadratic order fluctuated action in the fluctuations $\tilde{{g}}$ and $\tilde{\xi}$. Note that there is no linear term in \cref{203} as we consider the action is not coupled to any external sources of $\xi$.  Replacing \cref{203} into \cref{200} and ignoring the  higher order terms, we have
\begin{equation}\label{204}
	\text{ln}\thinspace Z \approx -\mathcal{S}[\bar{g}, \bar{\xi}]+ \text{ln}\int D[\tilde{g},\tilde{\xi}]\text{exp}(-\mathcal{S}_2[\tilde{{g}},\tilde{\xi}])
\end{equation}
$\mathcal{S}[\bar{g}, \bar{\xi}]$ is the background field contribution to the partition function while the second term of \cref{204} is regarded as the contribution of the matter quanta and thermal gravitons (fluctuation of metric) on the background geometry \cite{Hawking:1978td,Hawking:1977te}.
\subsection{The one-loop  effective action} 
The quantum corrected one-loop effective action $W$ \cite{Avramidi:1994th,Denardo:1982tb} is expressed as
\begin{equation}\label{205}
	W=-\text{ln}\thinspace \int D[\tilde{g},\tilde{\xi}]\text{exp}(-\mathcal{S}_2[\tilde{{g}},\tilde{\xi}])
\end{equation}
The quadratic term of action $\mathcal{S}_2$ in \cref{203} can be expressed as
\begin{equation}\label{206}
	\mathcal{S}_2[\tilde{{g}},\tilde{\xi}] = \int_{\mathcal{M}}d^4x \sqrt{\text{det}\thinspace \bar{g}}\thinspace \tilde{\xi}\Lambda \tilde{\xi},
\end{equation}
where $\Lambda$ is a differential operator which controls the quadratic order quantum fluctuations around the classical background. For bosonic fields,  $\Lambda$ is a second-order differential operator. Again, it is real, elliptic and self-adjoint since our background metric $\bar{g}$ has Euclidean signature (i.e.\ positive and real). But for fermionic fields, $\Lambda$ is of first-order. 
We will derive all the important results for bosonic fields first  and then extend them for fermionic fields.
If $\tilde{\xi}$ is a graviton or gauge field then $\Lambda$ will receive a complicated tensor structure, which has been suppressed here for simplicity. 

Considering $\Lambda$ has a discrete spectrum of eigenvalues $\lambda_i$, the \textit{one-loop effective action} given \cref{205} will take the following form \cite{Peixoto:2001wx}
\begin{equation}\label{207}
	W = \frac{\chi}{2}\thinspace \text{ln}\thinspace \text{det}\thinspace \Lambda = \frac{\chi}{2}\sum_i \text{ln}\thinspace \lambda_i,
\end{equation} 
where $\chi = \pm 1$ for bosons and fermions respectively. Note that the spectrum of $\Lambda$ may in practice also be continuous, and our basic structure of the heat kernel method and the general Seeley-DeWitt computation approach (\cref{ghkm,HKE}) are valid for both the discrete and continuous spectrums.

\subsection{The heat kernel method}\label{ghkm} 
Practically, one finds it difficult to compute $W$ directly, especially when fluctuations are gauge fields and gravitons. The most straightforward method is to compute the summation in \cref{207} provided we have the knowledge about the complete basis of normalized eigenfunctions $\lbrace f_i\rbrace$ and eigenvalues $\lbrace \lambda_i \rbrace$ obeying $\Lambda f_i = \lambda_i f_i$. But this is a challenging task. Also, from the series in \cref{207} it is evident that the one loop determinant det$\Lambda$ diverges (UV-divergence in quantum field theory) and traditional way to regulate this is the \textit{heat kernel method}: expressing $W$ in terms of diagonal matrix element of heat kernel via a special integral representation, called proper time representation \eqref{212}.

In terms of eigenfunctions $\lbrace f_i^{m}\rbrace$ and eigenvalues $\lbrace \lambda_i\rbrace$ of the kinetic operator $\Lambda$, \textit{heat kernel} (of the operator $e^{-s\Lambda}$) can be defined as \cite{Sen:2012rr,Sen:2012qq}
\begin{equation}\label{208}
	K^{mn}(x,y;s) = \sum_i e^{-\lambda_is}f_i^m(x)f_i^n(y),
\end{equation}
where $m$, $n$ are set of indices of the eigenfunctions at two points $x$, $y$ respectively on $\mathcal{M}$; $s$ is an auxiliary time variable, called proper time. The eigenfunctions  $\lbrace f_i^{m}\rbrace$ have been normalized so that
\begin{equation}\label{209}
	\sum_i G_{mn}f_i^{m}(x)f_i^{n}(y) = \delta^4(x,y), \enspace \int_{\mathcal{M}}d^4x \sqrt{\text{det}\thinspace \bar{g}}\thinspace G_{mn} f_i^{m}(x)f_j^{n}(x) = \delta_{ij},
\end{equation}
where $G_{mn}$ is a metric in the space of eigenfunctions (fields) induced by $\bar{g}$. Contracting over indices $m,n$, let us define the heat kernel $K(x,y;s)$
\begin{equation}\label{210}
	K(x,y;s) = G_{mn}K^{mn}(x,y;s),
\end{equation}
which satisfies the diffusion equation
\begin{equation}\label{215}
	\left(\frac{\partial}{\partial s}+\Lambda_x \right)K(x,y;s) = 0,
\end{equation}
with boundary conditions
\begin{equation}\label{216}
	K(x,y;0) = \delta (x,y),
\end{equation}
where $\Lambda_x$ describes the fact that in \cref{215} the operator $\Lambda$ is supposed to act only on the first argument of $K(x,y;s)$.
The trace of heat kernel $D(s)$, called \textit{heat trace}, can be obtained by setting $x=y$ (for avoiding the dependence of eigenfunctions) in \cref{210} and then integrating over all the $\mathcal{M}$:
\begin{equation}\label{211}
	D(s) = \int_{\mathcal{M}} d^4x \sqrt{\text{det}\thinspace \bar{g}}\thinspace K(x,x;s)= \sum_i e^{-\lambda_i s}.
\end{equation}
Heat trace $D(s)$ is important because it stores information about the spectrum of the kinetic operator $\Lambda$. For example, one may evaluate\footnote{In deriving \cref{212}, we have used the identity
	$\lim_{\epsilon\to 0}\int_\epsilon^\infty \frac{ds}{s}(e^{-As}-e^{-Bs}) = \text{ln} \thinspace\frac{B}{A}$}
\begin{equation}\label{212}
	W= -\frac{\chi}{2}\int_\epsilon^\infty \frac{ds}{s}\sum_i e^{-\lambda_i s}= -\frac{1}{2}\int_\epsilon^\infty \frac{ds}{s}\chi D(s).
\end{equation}
This is called Schwinger-DeWitt proper time representation \cite{Schwinger:1951sp,DeWitt:1975ps} of the effective action, where $\epsilon$ is an ultraviolet cutoff, naturally restricted by the Planck length, i.e.\ $\epsilon\sim {l_p}^2\sim G_N$. Also, one may identify the one-loop corrected effective Lagrangian density $\Delta\mathcal{L}_{\text{eff}}$ from \cref{212} as
\begin{equation}\label{214}
	\Delta\mathcal{L}_{\text{eff}}= -\frac{1}{2}\int_\epsilon^\infty \frac{ds}{s}\chi K(x,x;s).
\end{equation}
The main benefit of this approach lies in the fact that heat kernels are ultraviolet finite for all positive proper time ($s$).
\subsection{Heat kernel expansion and Seeley-DeWitt coefficients}
The conventional way of computing heat kernel is expanding them in powers of proper time $s$, called \textit{heat kernel expansion}. It is noticed that $K(x,x;s)$ is analytic for $s>0$, but as $s\to 0$ there is a power-law asymptotic expansion, known as Seeley-DeWitt expansion\footnote{$\chi$ has been absorbed in the definition \eqref{115} of $a_{2n}$ by manually putting proper signature following the spin-statistics.} \cite{Seeley:1966tt,Seeley:1969uu,DeWitt:1965ff,DeWitt:1967gg,DeWitt:1967hh,
	DeWitt:1967ii,Duff:1977vv,Christensen:1979ww,Christensen:1980xx,Duff:1980yy,Birrel:1982zz,Gilkey:1984xy,Duff:2011yz},
\begin{equation}\label{108}
	\chi K(x,x;s) \cong \sum_{n=0}^\infty s^{n-2}a_{2n}(x),
\end{equation}
giving
\begin{equation}
	\chi D(s)= \int_{\mathcal{M}} d^4x \sqrt{\text{det}\thinspace \bar{g}}\thinspace\left\lbrace\frac{1}{s^2}a_0(x) + \frac{1}{s}a_2(x) + a_4(x)+\cdots \right\rbrace,
\end{equation}
where the expansion coefficients $\lbrace a_{2n}(x)\rbrace$ are heat trace asymptotics, called \textit{Seeley-DeWitt coefficients} \cite{Seeley:1966tt,Seeley:1969uu,DeWitt:1965ff,DeWitt:1967gg,DeWitt:1967hh,DeWitt:1967ii} or heat kernel coefficients. A systematic way of computing these coefficients is reviewed in the next section, where  $\lbrace a_{2n}(x)\rbrace$ are expressed in terms of background fields and their covariant derivatives associated with the kinetic differential operator $\Lambda$.
\section{General approach for computing Seeley-DeWitt coefficients}\label{HKE}
The most common method for calculating Seeley-DeWitt coefficients in physics is the DeWitt iterative technique (reviewed in sec. 4.3 of \cite{Vassilevich:2003ll}). But, we will follow the method of Gilkey \cite{Gilkey:1975cd} (explained in sec. 4.1 of \cite{Vassilevich:2003ll})  which seems to be more efficient for a curved background with arbitrary spin. In this section, we will review the main ideas, the basic structure and useful formulae to compute the Seeley-DeWitt coefficients based on the  manual \cite{Vassilevich:2003ll}. We now start by considering the  quadratic order fluctuated action \eqref{206} controlled by the kinetic differential operator $\Lambda$ and rewrite its schematic form in terms of field indices as
\begin{equation}\label{110}
	\mathcal{S}_2 = \int d^4x \sqrt{\text{det}\thinspace \bar{g}}\thinspace \tilde{\xi}_m \Lambda^{mn}\tilde{\xi}_n,
\end{equation}
where \{$\tilde{\xi}_m$\} is the set of all fluctuating fields. In order to successfully apply the procedure, $\Lambda$ is constrained by the following conditions:
\begin{itemize}
	\item[\scalebox{0.6}{$\blacksquare$}] $\Lambda$ must be of a second order differential operator. It is useful to note that the quadratic order fluctuations of the fermions are governed by first order operators $\slashed{D}$, i.e.\ $\delta^2\mathcal{L} = \tilde{\xi} \slashed{D}\tilde{\xi}$. There we first restructure them as \textit{Dirac-type} and \textit{Hermitian} one, and define \cite{Sen:2012qq}
	\begin{equation}\label{116}
		\text{det}\thinspace \slashed{D} = \sqrt{\text{det}\slashed{D}^\dagger\slashed{D}}, \enspace \text{ln det}\thinspace \slashed{D} = \frac{1}{2} \text{ln det}\thinspace \Lambda,
	\end{equation}
	where $\Lambda=\slashed{D}^\dagger\slashed{D}=\slashed{D}^2$. This will include an additional $1/2$ factor in the description of heat kernel (or in the formula of Seeley-DeWitt coefficients), which is compensated by a factor of 2 arises due to the fact that squaring the complex operator doubles the integration variables. 
	\item[\scalebox{0.6}{$\blacksquare$}] $\Lambda$ must be made up of pseudo-differential operators which excludes negative and fractional powers of the derivatives.
	\item[\scalebox{0.6}{$\blacksquare$}] It should be elliptic and self-adjoint, i.e.\ $\langle \tilde{\xi}_m, \Lambda \tilde{\xi}_n\rangle = \langle \Lambda\tilde{\xi}_m,\tilde{\xi_n}\rangle$.
	\item[\scalebox{0.6}{$\blacksquare$}] $\Lambda$ should be in a \textit{minimal} form of type $g^{\mu\nu}D_\mu D_\nu+ Q$, where it has a scalar principal part, i.e.\ second order derivatives are contracted with the metric.
\end{itemize}
In summary, we conclude that the present method demands $\Lambda$ to be a \textit{Laplace-type}  operator having the following schematic form
\begin{equation}\label{111}
	\Lambda^{mn} = \pm\big\lbrace(D^\rho D_\rho)G^{mn} +(N^\rho D_\rho)^{mn}+P^{mn}\big\rbrace,
\end{equation}
where $G^{mn}$ is some arbitrary matrix,\footnote{Precisely, $G$ is the effective metric tensor which contracts indices of an operator acting on $\tilde{\xi}_m$ and we can also use $G$ to find trace values of these operators. For example, if $G^{mn}$ be the metric tensor related to an arbitrary two rank tensor $B^{mn}$, then  ${B^m}_n = B^{mp}G_{pn}$, $\text{tr}(B) = {B^m}_m$  and $\text{tr}(B^2) = {B^m}_n {B^n}_m$.} $D_\rho$ is the ordinary covariant derivative with 
connections determined by the background metric, and $N^\rho$, $P$ are arbitrary matrices constructed from the background fluctuations. Let us define a new covariant derivative $\mathcal{D}_\rho$  including the field connection $\omega_\rho$,
\begin{equation}\label{117}
	\mathcal{D}_\rho \tilde{\xi}_m = D_\rho \tilde{\xi}_m + {(\omega_\rho)_m}^n \tilde{\xi}_n \quad\forall m\neq n,
\end{equation}
and rewrite \cref{111} by replacing $D_\rho$ with $\mathcal{D}_\rho$ as 
\begin{equation}\label{112}
	\Lambda^{mn} = \pm\left\lbrace(\mathcal{D}^\rho\mathcal{D}_\rho)G^{mn}+P^{mn}-(D^\rho\omega_\rho)^{mn}-(\omega^\rho)^{mp}{(\omega_\rho)_p}^n+(N^\rho-2\omega^\rho)^{mn}D_\rho\right\rbrace.
\end{equation}
We aim to express \cref{112} in the standard form
\begin{equation}\label{113}
	\Lambda = \pm\big\lbrace(\mathcal{D}^\rho\mathcal{D}_\rho)I+E\big\rbrace,
\end{equation}
and extract the following necessary matrices,
\begin{align}\label{114}
	\begin{split}
		&I^{mn} = G^{mn}, (\omega_\rho)^{mn} = \frac{1}{2}(N_\rho)^{mn},\\
		& (E)^{mn} = P^{mn}-(D^\rho\omega_\rho)^{mn}-(\omega^\rho)^{mp}{(\omega_\rho)_p}^n,\\
		&(\Omega_{\rho\sigma})^{mn} = [D_\rho,D_\sigma]^{mn}+{D_{[\rho}\omega_{\sigma]}} ^{mn}+[\omega_\rho,\omega_\sigma]^{mn},
	\end{split}
\end{align}
where $\Omega_{\rho\sigma}\equiv [\mathcal{D}_\rho,\mathcal{D}_\sigma]$ is the curvature associated with $\mathcal{D}_\rho$. So, one needs to identify the matrices $G$, $P$ and $N$ from the relation \eqref{112} and use those results in the formula \eqref{114} for obtaining $\omega, E$ and $\Omega$. With these definitions, the formulae for
the first three Seeley-DeWitt coefficients \cite{Vassilevich:2003ll,Karan:2018ac}
are summarized as following
\begin{align}\label{115}
	\begin{split}
		(4\pi)^2a_0(x) &= \text{tr}(I),\\
		(4\pi)^2a_2(x) &=\frac{1}{6} \text{tr}(6E +RI),\\
		(4\pi)^2a_4(x) &= \frac{1}{360} \text{tr}\big( 60RE+ 180E^2 + 30\Omega_{\rho\sigma}\Omega^{\rho\sigma}+(5 R^2+ 2R_{\mu\nu\rho\sigma}R^{\mu\nu\rho\sigma}-2R_{\mu\nu}R^{\mu\nu})I\big).
	\end{split}
\end{align}
The formulae given in \cref{115} are applicable only for manifolds having no boundary.\footnote{There are Seeley-DeWitt coefficient formulae for manifold with boundary (see eqs. (5.29)-(5.33) of \cite{Vassilevich:2003ll}) also. But in this paper, our main motive is to compute logarithmic corrections of black hole entropy which demands to choose the no boundary formulae (discussed in \cref{app}).} Hence, all the 
total derivative terms have been ignored as
they will appear as boundary terms in the integral
\eqref{108} and hence will disappear after integration. Again for manifolds satisfying the condition $R=0$ one can neglect terms proportional to $R$ in this formulae.
We consider free\footnote{\enquote{Free} means we are considering only interaction with background geometry, all other interactions are turned off.} quantum fields on a Riemannian manifold having no boundaries so that the Seeley-DeWitt coefficients can be described by local invariants induced from the background fields and metric. 
The main advantage of calculating the Seeley-DeWitt coefficients following this general Seeley-DeWitt approach is — (a) after considering quadratic fluctuations for different fields in a given theory we have simple and straightforward formulae to compute the Seeley-DeWitt coefficients, (b) the coefficients are expressed in terms of geometric invariants, hence one can employ these results for any arbitrary background field configuration without imposing any limitation on the background geometry. 
It is important to mention that for the fermionic case, we need to multiply an additional minus sign (correspond to $\chi=-1$) in the above formulae of  Seeley-DeWitt coefficients. We also note that the present method assumes the fermionic field fluctuations as Dirac spinors, and hence for casting Majorana and Weyl (having half the degrees of freedom of Dirac spinor)  fermion contribution, one needs to multiply a factor of $-1/2$ in the formulae \eqref{115}. 

Finally, we summarize the key steps of computing Seeley-DeWitt coefficients as follows:\footnote{For a better understanding of this method and some elementary examples, readers may refer \cite{Karan:2018ac,Charles:2015nn}.}
\begin{enumerate}
	\item Consider fluctuations for all possible fields present in the theory around a classical background and express the action in quadratic order of fluctuations.
	\item Gauge fix the theory of choice by adding appropriate gauge fixing term to the quadratic action and find the quadratic fluctuation controlling operator $\Lambda$ (defined in the \cref{110}). Identify the appropriate ghosts and keep them separate. 
	\item With all the restrictions mentioned, restructure $\Lambda$ so that we can have the canonical form \eqref{111}. Extract the results for the matrices $G$, $P$ and $N$.
	\item Define a new effective covariant derivative $\mathcal{D}_\rho$ with field connection $\omega$ as \cref{117} and making use of it, modify $\Lambda$ further to have the form \eqref{112}.
	\item Comparing the new form of $\Lambda$ with standard form \eqref{113} read off the results for matrices $I,E$ and $\Omega$ using \cref{114}.
	\item Compute $\text{tr}\thinspace (I)$, $\text{tr}\thinspace (E)$, $\text{tr}\thinspace (E^2)$ and $\text{tr}\thinspace (\Omega^2)$. Use this data in the formulae \eqref{115} to find the result for the first three Seeley-DeWitt coefficients (without ghosts) $a_{2n}^{\text{no ghost}}$. 
	\item Making use of background equations of motion, gamma and tensor identities, simplify $a_{2n}^{\text{no ghost}}$.
	\item Repeat the same steps ($3-7$) starting from ghost action to have Seeley-DeWitt coefficients for ghosts ($a_{2n}^{\text{ghost}}$).
	\item  Obtain the total Seeley-DeWitt coefficients of the theory by $a_{2n}^{\text{total}}= a_{2n}^{\text{no ghost}}+ a_{2n}^{\text{ghost}}$.
\end{enumerate}

\section{Minimal $\mathcal{N}=2, d=4$ Einstein-Maxwell supergravity theory}\label{EMSGT}
In this section, our motive is to briefly summarize a minimally coupled $\mathcal{N}=2$ supergravity embedded with the solutions of Einstein-Maxwell theory in four dimensions.
There are a class of $\mathcal{N}=2$ supergravity solutions to the equations of motion, determined by certain conditions \cite{Charles:2017aa}. These conditions reduce the full $\mathcal{N}=2$ supergravity equations of motion to a much simpler form of equations of motion for Einstein-Maxwell theory. The resultant minimal $\mathcal{N}=2, d=4$ Einstein-Maxwell supergravity theory (EMSGT) plays a significant role in string phenomenology, black holes, and holography.
In general, $\mathcal{N}=2$ supergravity theories are  coupled to additional matter fields like $n_V$ vector multiplets and $n_H$ hyper multiplets \cite{Itoyama:2006wp,Charles:2017aa}. But in this work, we will focus on the simpler case of  minimally coupled $\mathcal{N}=2$ supergravity where $n_V=n_H=0$. However, both the theory have the same type of classical or black hole solutions. At any point, one can compute the Seeley-DeWitt coefficients and the logarithmic contributions for $n_V$ and $n_H$ fluctuations separately, and can extend the minimal supergravity results to these matter coupled theories. The minimal $\mathcal{N}=2$ EMSGT is associated with the field content provided in table \ref{tab1}. Both the bosonic and fermionic sectors of the field content are described by the Lagrangians $\mathcal{L}_{\text{EM}}$ and $\mathcal{L}_{\text{gravitini}}$ respectively in the action \eqref{EM}. Note that all these fields are free in the sense that they are only coupled to the metric (or graviton).
{
	\renewcommand{\arraystretch}{1.5}
	\begin{table}[t]
		\centering
		\hspace{-0.2in}
		\begin{tabular}{|>{\centering}p{4.2in}|}
			\hline
			\textbf{Physical Field Content}  \tabularnewline \hline
			Bosonic sector: one graviton ($g_{\mu\nu}$), one graviphoton ($A_{\mu}$) \tabularnewline
			Fermionic sector: two gravitino ($\psi_{\mu}, \phi_{\mu}$) \tabularnewline
			($\mu,\nu = 1 \cdots 4$) \tabularnewline \hline
		\end{tabular}
		\caption{Field content in minimally coupled $\mathcal{N}=2$ EMSGT.}\label{tab1}
	\end{table}
} 

We consider standard $d=4$ Einstein-Maxwell theory embedded in minimal $\mathcal{N}=2$ supergravity described by the following action,
\begin{equation}\label{EM}
	\mathcal{S} =  \int d^4x \sqrt{\text{det}\thinspace {g}} \thinspace (\mathcal{L}_{\text{EM}} + \mathcal{L}_{\text{gravitini}} ), 
\end{equation}
with
\begin{align}\label{EM1}
	\begin{split}
		&\mathcal{L}_{\text{EM}} = \left(\mathcal{R}-F_{\mu\nu}F^{\mu\nu} \right),\\
		&\mathcal{L}_{\text{gravitini}} = -\frac{1}{2}\bar{\psi}_{\mu}\gamma^{\mu \rho\nu} D_\rho\psi_{\nu}-\frac{1}{2}\bar{\phi}_{\mu}\gamma^{\mu \rho\nu} D_\rho\phi_{\nu}\\
		&\qquad\qquad\quad+\frac{1}{2}\bar{\psi}_{\mu}\big({F}^{\mu\nu}+\frac{1}{2}\gamma^5 {H}^{\mu\nu} \big)\phi_{\nu}- \frac{1}{2}\bar{\phi}_{\mu}\big({F}^{\mu\nu}+\frac{1}{2}\gamma^5 {H}^{\mu\nu} \big)\psi_{\nu},
	\end{split}
\end{align}
where $\mathcal{R}$ is the Ricci scalar induced from metric $g_{\mu\nu}$, $F_{\mu\nu}$ the electromagnetic field strength of gauge field $A_\mu$ in Einstein-Maxwell theory, $\psi_\mu,\phi_\mu$ are two gravitini species
and ${H}^{\mu\nu}=-\varepsilon^{\mu\nu\rho\sigma}{F}_{\rho\sigma} $  is the dual field strength.\footnote{In particular, the identity $\gamma^{\mu\nu\rho\sigma}=-\gamma^5\varepsilon^{\mu\nu\rho\sigma}$ is very much useful, where $\gamma^5 = -\frac{1}{4!}\varepsilon^{\mu\nu\rho\sigma}\gamma_\mu\gamma_\nu\gamma_\rho\gamma_\sigma$ in four dimensions.}  $\psi_\mu,\phi_\mu$ are Majorana spinors in 4D and the action \eqref{EM} describes black holes in pure $\mathcal{N}=2$ supergravity. It is useful to note that we have set\footnote{We are following the structure and conventions of \cite{Bhattacharyya:2012ss}.} $G_N=1/{16\pi}$.
\subsection{Equations of motion and identities}\label{EOM}
Let us denote $(\bar{g}_{\mu\nu}, \bar{A}_\mu)$ as an arbitrary solution to the classical equations of motion of our theory and define corresponding background field strength tensor $\bar{F}_{\mu\nu}\equiv \partial_\mu \bar{A}_\nu-\partial_\nu \bar{A}_\mu$. Then the geometry and matter fields are related by Einstein equation, 
\begin{equation}\label{Einstein}
	R_{\mu\nu}-\frac{1}{2}\bar{g}_{\mu\nu}R= 2\bar{F}_{\mu\rho}{\bar{F_\nu}}^\rho-\frac{1}{2}\bar{g}_{\mu\nu}\bar{F}_{\rho\sigma}\bar{F}^{\rho\sigma}.
\end{equation} 
The background gauge field strength $\bar{F}_{\mu\nu}$ satisfies Maxwell equations and Bianchi identity,
\begin{equation}\label{Maxwell}
	D^\mu \bar{F}_{\mu\nu} =0,\\\\ \enspace D_{[\mu}\bar{F}_{\nu\rho]}=0.
\end{equation}
Eq.(\ref{Einstein}) and eq.(\ref{Maxwell}) represents
the basic equations of motion of $\mathcal{N}=2, d=4$ EMSGT. Again, the trace of Eq.(\ref{Einstein}) yields
\begin{equation}\label{116}
	R=0,
\end{equation}
Hence for the remaining parts of this paper, we will use this condition and ignore the terms proportional to  $R$. Again using \eqref{116}, \eqref{Einstein} reduces to,
\begin{equation}\label{keyIden}
	\bar{F}_{\mu\rho}{\bar{F_\nu}}^\rho = \frac{1}{2}R_{\mu\nu}+\frac{1}{4}\bar{g}_{\mu\nu}\bar{F}_{\rho\sigma}\bar{F}^{\rho\sigma}.
\end{equation} 
In addition, the dual gauge field strength $\bar{H}_{\mu\nu}$ satisfies the Maxwell-Bianchi equations of the form,
\begin{equation}\label{MB}
	D^\mu \bar{H}_{\mu\nu} =0, \enspace D_{[\mu}\bar{H}_{\nu\rho]}=0.
\end{equation}
It is also important to mention the gravitational Bianchi identity
\begin{equation}\label{gravbianchi}
	R_{\mu[\nu\rho\sigma]}=0.
\end{equation}
We have used \cref{Maxwell,keyIden,MB,gravbianchi} to simplify our calculations and express them in closed forms.
\section{Seeley-DeWitt coefficients in minimal $\mathcal{N}=2, d=4$ Einstein-Maxwell supergravity theory}\label{HKIEMSGT}
In this section, we will calculate the first three Seeley-DeWitt coefficients of the quadratic order fluctuations of minimally coupled $\mathcal{N}=2, d=4$ EMSGT fields  discussed in \cref{EMSGT}. The field content is divided into two sectors — fermions and bosons, for which the contributions will be calculated  separately. We actually compute results for the fermionic sector, while the bosonic sector results are reviewed and recalculated from \cite{Bhattacharyya:2012ss}.
Finally, we will sum up these results in order to find the complete Seeley-DeWitt coefficients of the theory.


\subsection{$\mathcal{N}=2$ SUGRA: Bosons}\label{bosons}
Let's now consider the Einstein-Maxwell theory described by the Lagrangian $ \mathcal{L}_{\text{EM}}$ in \eqref{EM}  with the action,

\begin{equation}\label{32}
	\mathcal{S}_{\text{EM}} = \int d^4x \sqrt{\text{det}\thinspace {g}}\thinspace(\mathcal{R}-F_{\mu\nu}F^{\mu\nu}).
\end{equation}
We aim to calculate the  Seeley-DeWitt coefficients for this theory by considering quadratic order fluctuations around a classical background. This methodology has been developed in \cite{Bhattacharyya:2012ss}, where the authors have computed only the  $a_4$ coefficient  for calculating the logarithmic correction to extremal Kerr-Newman black hole entropy. But we find it worthful to briefly review this work and calculate all the first three Seeley-DeWitt coefficients for the theory, which may be useful in other applications. 

Let  $(\bar{g}_{\mu\nu}, \bar{A}_\alpha)$  be any arbitrary classical solution to the equations of motion of Einstein-Maxwell theory, and around this classical background we consider following fluctuations of metric and gauge field\footnote{The graviton and graviphoton have been scaled in such a manner that the kinetic terms for them have the same normalization in the quadratic order fluctuated action \eqref{34}.}
\begin{equation}\label{33}
	\tilde{g}_{\mu\nu} = \sqrt{2}h_{\mu\nu},\enspace  \tilde{A}_\alpha = \frac{1}{2} a_\alpha,
\end{equation}
where $h_{\mu\nu}$ and $a_\alpha$ are the graviton and graviphoton respectively. Moreover, the theory is gauge fixed by adding the following gauge-fixing term,
\begin{align}
	\begin{split}
		& -\int d^4x \sqrt{\text{det}\thinspace \bar{g}}\thinspace \Big\lbrace ( D^\mu h_{\mu\rho}-\frac{1}{2} D_\rho h)(D^\nu {h_\nu}^\rho-\frac{1}{2}D^\rho h)+\frac{1}{2} (D^\mu a_\mu)(D^\nu a_\nu)\Big\rbrace
	\end{split}
\end{align}
and the corresponding ghost action is added to the action \eqref{32}.
Up to  quadratic order in the fluctuations the gauge fixed action  excluding the ghost part is given by\cite{Bhattacharyya:2012ss},

{
	\allowdisplaybreaks
	\begin{align}\label{34}
		\begin{split}
			&\mathcal{S}_2 = \frac{1}{2}\int d^4x \sqrt{\text{det}\thinspace \bar{g}}\thinspace \tilde{\xi}_m\Lambda^{mn}\tilde{\xi}_n,\\
			&\tilde{\xi}_m\Lambda^{mn}\tilde{\xi}_n = h_{\mu\nu}\Box h^{\mu\nu} -\frac{1}{2}h\Box h +a_\alpha\Box a^\alpha -a_\alpha R^{\alpha\beta} a_\beta\\
			&\qquad\qquad\quad + h_{\mu\nu}\Big\lbrace 2R^{\mu\alpha\nu\beta}+\bar{g}^{\mu\alpha} R^{\nu\beta} +\bar{g}^{\alpha\beta}R^{\mu\nu}-\bar{g}^{\mu\nu}R^{\alpha\beta}  -3\bar{g}^{\nu\beta} R^{\mu\alpha} \\
			& \qquad\qquad\quad -4\bar{F}^{\mu\alpha} \bar{F}^{\nu\beta}+ (\frac{1}{2}\bar{g}^{\mu\nu}\bar{g}^{\alpha\beta}-\bar{g}^{\mu\alpha} \bar{g}^{\nu\beta})(\bar{F}_{\theta\varphi}\bar{F}^{\theta\varphi})\Big\rbrace h_{\alpha\beta}\\
			&\qquad\qquad\quad +h_{\mu\nu}\left\lbrace 4\sqrt{2}\bar{g}^{\alpha\nu} \bar{F}^{\rho\mu}- 4\sqrt{2}\bar{g}^{\nu\rho} \bar{F}^{\alpha\mu}-2\sqrt{2}\bar{g}^{\mu\nu}\bar{F}^{\rho\alpha}\right\rbrace (D_\rho a_\alpha),
		\end{split}
	\end{align}
}
where we denote $\Box \equiv D_\mu D^\mu$, $h={h^\mu}_\mu=\text{tr}(h_{\mu\nu})$, $\bar{F}_{\mu\nu}\equiv \partial_\mu \bar{A}_\nu-\partial_\nu \bar{A}_\mu$ and $\tilde{\xi}_m = \lbrace h_{\mu\nu}, a_\alpha \rbrace$.
Kindly note that $\Lambda$ is not in Laplace-type form due to the presence of the kinetic term $h\Box h$. We need to restructure $\Lambda$ and express it in the canonical form. In order to proceed further, we introduce the operator $G^{\mu\nu\thinspace\alpha\beta}$\footnote{The general form of $G$ is $\tilde{\xi}_m G^{mn}\tilde{\xi}_n = h_{\mu\nu}G^{\mu\nu\thinspace\alpha\beta}h_{\alpha\beta}+  a_\alpha G^{\alpha\beta}a_\beta$, where the form of $G^{\mu\nu\thinspace\alpha\beta}$ is given in \cref{35} and $G^{\alpha\beta} = \bar{g}^{\alpha\beta}$.}
\begin{equation}\label{35}
	G^{\mu\nu\thinspace\alpha\beta} = \frac{1}{2}(\bar{g}^{\mu\alpha} \bar{g}^{\nu\beta} + \bar{g}^{\mu\beta} \bar{g}^{\nu\alpha}-\bar{g}^{\mu\nu}\bar{g}^{\alpha\beta}),
\end{equation}
and rewrite the kinetic part as $h_{\mu\nu}\Box h^{\mu\nu} -\frac{1}{2}h\Box h +a_\alpha\Box a^\alpha = G^{\mu\nu\thinspace\alpha \beta}h_{\mu\nu}\Box h_{\alpha\beta}+ \bar{g}^{\alpha\beta} a_\alpha\Box a_\beta $. 
With necessary modifications of the total derivative terms to make $\Lambda$ Hermitian, the resulting form of $\Lambda$ becomes
{
	\allowdisplaybreaks
	\begin{align}\label{36}
		\begin{split}
			\tilde{\xi}_m\Lambda^{mn}\tilde{\xi}_n &= G^{\mu\nu\thinspace\alpha\beta}h_{\mu\nu}\Box h_{\alpha\beta}+ \bar{g}^{\alpha\beta} a_\alpha\Box a_\beta-a_\alpha R^{\alpha\beta}a_\beta\\
			&\quad + h_{\mu\nu}\Big\lbrace  R^{\mu\alpha\nu\beta}+R^{\mu\beta\nu\alpha}-\frac{1}{2}(\bar{g}^{\mu\alpha}R^{\nu\beta}+ \bar{g}^{\mu\beta}R^{\nu\alpha}\\
			&\quad +\bar{g}^{\nu\alpha}R^{\mu\beta}+\bar{g}^{\nu\beta}R^{\mu\alpha})-2(\bar{F}^{\mu\alpha}\bar{F}^{\nu\beta}+\bar{F}^{\mu\beta}\bar{F}^{\nu\alpha})\\
			&\quad+\frac{1}{2}(\bar{g}^{\mu\nu}\bar{g}^{\alpha\beta}-\bar{g}^{\mu\alpha}\bar{g}^{\nu\beta}-\bar{g}^{\mu\beta}\bar{g}^{\nu\alpha})(\bar{F}_{\theta\varphi}\bar{F}^{\theta\varphi})\Big\rbrace h_{\alpha\beta}\\
			&\quad -h_{\mu\nu}\Big\lbrace \frac{1}{4}(D_\rho K^\rho)^{\mu\nu\thinspace \alpha}-\frac{1}{2}(K^\rho)^{\mu\nu\thinspace \alpha}D_\rho\Big\rbrace a_\alpha\\
			&\quad -a_\alpha\Big\lbrace \frac{1}{4}(D_\rho K^\rho)^{\mu\nu\thinspace \alpha}+\frac{1}{2}(K^\rho)^{\mu\nu\thinspace \alpha}D_\rho\Big\rbrace h_{\mu\nu},
		\end{split}
	\end{align}
}
where 
\begin{equation}\label{37}
	(K^\rho)^{\mu\nu\thinspace \alpha} = 2\sqrt{2}(\bar{g}^{\alpha\mu}\bar{F}^{\rho\nu}+\bar{g}^{\alpha\nu}\bar{F}^{\rho\mu}-\bar{g}^{\mu\rho}\bar{F}^{\alpha\nu}-\bar{g}^{\nu\rho}\bar{F}^{\alpha\mu}-\bar{g}^{\mu\nu}\bar{F}^{\rho\alpha}).
\end{equation}
We identify the two matrices $P$ and $N$ (defined in \eqref{111}) as
{
	\allowdisplaybreaks
	\begin{align}\label{38}
		\begin{split}
			&\tilde{\xi}_m {(N^\rho)}^{mn}\tilde{\xi}_n = \frac{1}{2}h_{\mu\nu}(K^\rho)^{\mu\nu\thinspace \alpha}a_\alpha-\frac{1}{2}a_\alpha(K^\rho)^{\mu\nu\thinspace \alpha}h_{\mu\nu},\\
			&\tilde{\xi_m} P^{mn}\tilde{\xi_n} = h_{\mu\nu}\Big\lbrace  R^{\mu\alpha\nu\beta}+R^{\mu\beta\nu\alpha}-\frac{1}{2}(\bar{g}^{\mu\alpha}R^{\nu\beta}+ \bar{g}^{\mu\beta}R^{\nu\alpha}\\
			&\qquad\qquad\quad +\bar{g}^{\nu\alpha}R^{\mu\beta}+\bar{g}^{\nu\beta}R^{\mu\alpha})-2(\bar{F}^{\mu\alpha}\bar{F}^{\nu\beta}+\bar{F}^{\mu\beta}\bar{F}^{\nu\alpha})\\
			&\qquad\qquad\quad+\frac{1}{2}(\bar{g}^{\mu\nu}g^{\alpha\beta}-\bar{g}^{\mu\alpha}g^{\nu\beta}-\bar{g}^{\mu\beta}\bar{g}^{\nu\alpha})(\bar{F}_{\theta\varphi}\bar{F}^{\theta\varphi})\Big\rbrace h_{\alpha\beta}\\
			&\qquad\qquad\quad -a_\alpha R^{\alpha\beta}a_\beta+\frac{\sqrt{2}}{2}h_{\mu\nu}\Big\lbrace D^\mu \bar{F}^{\alpha\nu}+ D^\nu \bar{F}^{\alpha\mu}\Big\rbrace a_\alpha\\
			&\qquad\qquad\quad +\frac{\sqrt{2}}{2}a_\alpha\Big\lbrace D^\mu \bar{F}^{\alpha\nu}+ D^\nu \bar{F}^{\alpha\mu}\Big\rbrace h_{\mu\nu}.
		\end{split}
	\end{align} 
}
By comparing $\Lambda$ from \cref{36} with the standard form \eqref{113}, one can  extract the useful matrices $I$, $\omega$, $E$ and $\Omega$, and find their traces in order to compute the Seeley-DeWitt coefficients.
It is now straightforward to express the identity matrix as 
\begin{equation}\label{39}
	\tilde{\xi_m} I^{mn}\tilde{\xi_n} = h_{\mu\nu} G^{\mu\nu\thinspace\alpha\beta}h_{\alpha\beta}+ a_\alpha \bar{g}^{\alpha\beta}a_\beta.
\end{equation}
Making use of the results of $N$ and $P$ from \cref{38} in the formulae \eqref{114}, we express
{
	\allowdisplaybreaks
	\begin{align}\label{40}
		\begin{split}
			&\tilde{\xi}_m (\omega^\rho)^{mn}\tilde{\xi}_n =  \frac{1}{4}h_{\mu\nu}(K^\rho)^{\mu\nu\thinspace \alpha}a_\alpha-\frac{1}{4}a_\alpha(K^\rho)^{\mu\nu\thinspace \alpha}h_{\mu\nu},\\
			&\tilde{\xi}_m E^{mn}\tilde{\xi}_n = h_{\mu\nu} (R^{\mu\alpha\nu\beta}+ R^{\mu\beta\nu\alpha}-\bar{g}^{\mu\nu}R^{\alpha\beta}-\bar{g}^{\alpha\beta}R^{\mu\nu})h_{\alpha\beta}\\
			&\qquad\qquad\quad +\frac{3}{2}a_\alpha \bar{g}^{\alpha\beta}\bar{F}_{\mu\nu}\bar{F}^{\mu\nu}a_\beta+\frac{\sqrt{2}}{2}h_{\mu\nu}(D^\mu \bar{F}^{\alpha\nu}+ D^\nu \bar{F}^{\alpha\mu})a_\alpha\\
			&\qquad\qquad\quad +\frac{\sqrt{2}}{2}a_\alpha(D^\mu \bar{F}^{\alpha\nu}+ D^\nu \bar{F}^{\alpha\mu})h_{\mu\nu},\\
			&\tilde{\xi}_m(\Omega^{\rho\sigma})^{mn}\tilde{\xi}_n = h_{\mu\nu}\Big\lbrace\frac{1}{2}(\bar{g}^{\nu\beta}R^{\mu\alpha\rho\sigma}+\bar{g}^{\nu\alpha}R^{\mu\beta\rho\sigma}+\bar{g}^{\mu\beta}R^{\nu\alpha\rho\sigma}+\bar{g}^{\mu\alpha}R^{\nu\beta\rho\sigma})\\
			&\qquad\qquad\qquad\quad +[\omega^\rho,\omega^\sigma]^{\mu\nu\thinspace\alpha\beta}\Big\rbrace h_{\alpha\beta}+ a_\alpha\Big\lbrace R^{\alpha\beta\rho\sigma}+ [\omega^\rho,\omega^\sigma]^{\alpha\beta}\Big\rbrace a_\beta\\
			&\qquad\qquad\qquad\quad +h_{\mu\nu}{D^{[\rho}\omega^{\sigma]}}^{\mu\nu\thinspace\alpha}a_\alpha+ a_\alpha{D^{[\rho}\omega^{\sigma]}}^{\alpha\thinspace\mu\nu}h_{\mu\nu}
		\end{split}
	\end{align}
}
Note that in writing the expression for $\Omega$, we have used the following commutation relation,
\begin{align}\label{41}
	\begin{split}
		\tilde{\xi}_m [D^\rho,D^\sigma]^{mn}\tilde{\xi}_n &= h_{\mu\nu}[D^\rho,D^\sigma]h^{\mu\nu}+ a_\alpha [D^\rho,D^\sigma]a^\alpha\\
		&= h_{\mu\nu}\Big\lbrace \bar{g}^{\nu\beta}R^{\mu\alpha\rho\sigma}+ \bar{g}^{\mu\beta}R^{\nu\alpha\rho\sigma}\Big\rbrace h_{\alpha\beta}+ a_\alpha R^{\alpha\beta\rho\sigma}a_\beta.
	\end{split}
\end{align}
The useful traces for the matrices are finally written down in \eqref{42}. We skip the details of the intermediate steps as the calculation is straightforward although tedious.
\begin{align}\label{42}
	\begin{split}
		&\text{tr}(I) = 10+4=14,\\
		&\text{tr}(E) = 6\bar{F}_{\mu\nu}\bar{F}^{\mu\nu},\\
		&\text{tr}(E^2) = 3R_{\mu\nu\rho\sigma}R^{\mu\nu\rho\sigma}-7R_{\mu\nu}R^{\mu\nu}+3R_{\mu\nu\rho\sigma}\bar{F}^{\mu\nu}\bar{F}^{\rho\sigma}+9(\bar{F}_{\mu\nu}\bar{F}^{\mu\nu})^2,\\
		&\text{tr}(\Omega_{\rho\sigma}\Omega^{\rho\sigma})  = -7R_{\mu\nu\rho\sigma}R^{\mu\nu\rho\sigma}+56R_{\mu\nu}R^{\mu\nu}-18R_{\mu\nu\rho\sigma}\bar{F}^{\mu\nu}\bar{F}^{\rho\sigma}\\
		&\hspace{1in}-54(\bar{F}_{\mu\nu}\bar{F}^{\mu\nu})^2.
	\end{split}
\end{align}
Substituting these trace values in the formulae \eqref{115}, we find the Seeley-DeWitt coefficients of Einstein-Maxwell theory (without ghost). The results are
\begin{align}\label{43}
	\begin{split}
		&(4\pi)^2 a_0^{\text{EM}} = 14,\\
		&(4\pi)^2 a_2^{\text{EM}} = 6\bar{F}_{\mu\nu}\bar{F}^{\mu\nu},\\
		&(4\pi)^2 a_4^{\text{EM}} = \frac{1}{180}\left(179 R_{\mu\nu\rho\sigma}R^{\mu\nu\rho\sigma}+196 R_{\mu\nu}R^{\mu\nu} \right) .
	\end{split}
\end{align}
The ghost fields
are non-interacting and hence can be treated separately. The corresponding ghost action is given by,
\begin{equation}\label{44}
	\mathcal{S}_{\text{ghost},b} = \frac{1}{2}\int d^4x \sqrt{\text{det}\thinspace \bar{g}}\thinspace\left\lbrace 2b_\mu (\bar{g}^{\mu\nu}\Box+ R^{\mu\nu})c_\nu+2b\Box c-4b \bar{F}^{\rho\nu}D_\rho c_\nu\right\rbrace,
\end{equation}
where $b_{\mu},c_\mu$ (vector fields) are diffeomorphism ghosts related to the graviton; $b,c$ (scalar fields) are the ghosts associated with the graviphoton. Adjusting this action up to total derivative, we express the corresponding $\Lambda$ as, 
\begin{align}\label{45}
	\begin{split}
		\tilde{\xi}_m \Lambda^{mn}\tilde{\xi}_n &= b_\mu g^{\mu\nu}\Box c_\nu+c_\mu g^{\mu\nu}\Box b_\nu+b\Box c+c\Box b\\
		&\quad +b_\mu R^{\mu\nu} c_\nu+ c_\mu R^{\mu\nu} b_\nu-2b \bar{F}^{\rho\nu}D_\rho c_\nu + 2c_\mu \bar{F}^{\rho\mu}D_\rho b,
	\end{split}
\end{align}
and comparing with \eqref{111} identify the matrices $N$ and $P$ as
\begin{align}\label{46}
	\begin{split}
		&\tilde{\xi}_m (N^\rho)^{mn}\tilde{\xi}_n = -2b \bar{F}^{\rho\nu}c_\nu + 2c_\mu \bar{F}^{\rho\mu} b,\\
		& \tilde{\xi}_m P^{mn}\tilde{\xi}_n = b_\mu R^{\mu\nu}c_\nu + c_\mu R^{\mu\nu}b_\nu.
	\end{split}
\end{align}
We follow the  general approach  described in \cref{HKE} and making use of formulae \eqref{114} we extract the following necessary matrices
\begin{align}\label{47}
	\begin{split}
		&\tilde{\xi}_m I^{mn}\tilde{\xi}_n = b_\mu \bar{g}^{\mu\nu}c_\nu+ c_\mu \bar{g}^{\mu\nu}b_\nu+bc+cb,\\
		& \tilde{\xi}_m (\omega_\rho)^{mn}\tilde{\xi}_n = -b {\bar{F_\rho}}^\nu c_\nu + c_\mu \bar{F_\rho}^\mu b,\\
		& \tilde{\xi}_m E^{mn}\tilde{\xi}_n = b_\mu R^{\mu\nu}c_\nu + c_\mu R^{\mu\nu}b_\nu,\\
		& \tilde{\xi}_m (\Omega_{\rho\sigma})^{mn}\tilde{\xi}_n = b_\mu {R^{\mu\nu}}_{\rho\sigma}c_\nu+ c_\mu {R^{\mu\nu}}_{\rho\sigma}b_\nu+ b (D^\nu \bar{F}_{\rho\sigma}) c_\nu-c_\mu (D^\mu \bar{F}_{\rho\sigma})b.
	\end{split}
\end{align}
Using \eqref{47}, one can compute the following traces for the ghost fields, 
\begin{align}\label{49}
	\begin{split}
		&\text{tr}(I) = 4+4+1+1=10,\\
		&\text{tr}(E) = 0,\enspace\text{tr}(E^2) = 2R_{\mu\nu}R^{\mu\nu}, \\
		& \text{tr}(\Omega_{\rho\sigma}\Omega^{\rho\sigma})=-2R_{\mu\nu\rho\sigma}R^{\mu\nu\rho\sigma},
	\end{split}
\end{align} 
and can use them in order to find Seeley-DeWitt coefficients for ghost fields corresponding to the bosonic action of the theory. This yields
\begin{align}\label{50}
	\begin{split}
		&(4\pi)^2 a_0^{\text{ghost},b} = -10,\\
		&(4\pi)^2 a_2^{\text{ghost},b} = 0,\\
		&(4\pi)^2 a_4^{\text{ghost},b} =\frac{1}{18}\left(2 R_{\mu\nu\rho\sigma}R^{\mu\nu\rho\sigma}-17 R_{\mu\nu}R^{\mu\nu} \right).
	\end{split}
\end{align} 
Note that an overall minus sign is put  on each coefficient since ghosts fields follow the reverse of usual spin-statistics.
\subsection{$\mathcal{N}=2$ SUGRA: Fermions}\label{Fermions}
The fermionic sector of $\mathcal{N}=2$ includes two  Majorana gravitini fields described by the Lagrangian $ \mathcal{L}_{\text{gravitini}}$ expressed in \eqref{EM1}. To gauge fix the fermionic action we add the following term to  $ \mathcal{L}_{\text{gravitini}}$,

\begin{equation}
	\mathcal{L}_{\text{gf}} =\frac{1}{4}\bar{\psi}_{\mu}\gamma^{\mu}\gamma^{\rho}D_{\rho}\gamma^{\nu}\psi_{\nu}+\frac{1}{4}\bar{\phi}_{\mu}\gamma^{\mu}\gamma^{\rho}D_{\rho}\gamma^{\nu}\phi_{\nu}
\end{equation}
and we can express the quadratic part of the corresponding gauge fixed action in Dirac form as
{
	\allowdisplaybreaks
	\begin{equation}\label{11}
		\begin{gathered}
			\mathcal{S}_{{f}} = \int d^4x \sqrt{\text{det}\thinspace \bar{g}} \thinspace \mathcal{L}_{{f}},\\\mathcal{L}_{{f}} =
			\mathcal{L}_{\text{gravitini}} +\mathcal{L}_{\text{gf}} = \frac{i}{4}\bar{\psi}_{\mu}\gamma^\nu\gamma^\rho\gamma^\mu D_\rho\psi_{\nu}+\frac{i}{4}\bar{\phi}_{\mu}\gamma^\nu\gamma^\rho\gamma^\mu D_\rho\phi_{\nu}\\
			\hspace{1.8in}+\frac{i}{2}\bar{\psi}_{\mu}\big(\bar{F}^{\mu\nu}+\frac{1}{2}\gamma^5 \bar{H}^{\mu\nu} \big)\phi_{\nu}- \frac{i}{2}\bar{\phi}_{\mu}\big(\bar{F}^{\mu\nu}+\frac{1}{2}\gamma^5 \bar{H}^{\mu\nu} \big)\psi_{\nu},
		\end{gathered}
	\end{equation}
}
Now, we can rewrite the action for gravitini fluctuations $\tilde{\xi}_m = \lbrace\psi_\mu,\phi_\nu\rbrace$ as 
{
	\allowdisplaybreaks
	\begin{equation}\label{13}
		\begin{gathered}
			\mathcal{S}_2 = \frac{1}{2}\int d^4x \sqrt{\text{det}\thinspace \bar{g}} \thinspace \tilde{\xi}_m\slashed{D}^{mn} \tilde{\xi}_n,\\
			\tilde{\xi}_m\slashed{D}^{mn} \tilde{\xi}_n = \frac{i}{2}\bar{\psi}_{\mu}\gamma^\nu\gamma^\rho\gamma^\mu D_\rho\psi_{\nu}+\frac{i}{2}\bar{\phi}_{\mu}\gamma^\nu\gamma^\rho\gamma^\mu D_\rho\phi_{\nu}\\
			\hspace{1.8in}+i\bar{\psi}_{\mu}\big(\bar{F}^{\mu\nu}+\frac{1}{2}\gamma^5 \bar{H}^{\mu\nu} \big)\phi_{\nu}- i\bar{\phi}_{\mu}\big(\bar{F}^{\mu\nu}+\frac{1}{2}\gamma^5 \bar{H}^{\mu\nu} \big)\psi_{\nu}.
		\end{gathered}
	\end{equation}
}
The action contains first-order Dirac-type differential operator $\slashed{D}$, but for computing  the Seeley-DeWitt coefficients following general approach \cref{HKE}  we need to make it quadratic in order. With our choice of four-dimensional Euclidean space-time, gamma matrices are Hermitian, the gravitino conjugates are $\bar{\psi}_{\mu}=\psi_{\mu}^\dag, \bar{\phi}_{\mu}=\phi_{\mu}^\dag$, and $\left(\gamma^\nu\gamma^\rho\gamma^\mu D_\rho \right)^\dag= -\gamma^\nu\gamma^\rho\gamma^\mu D_\rho $. We will consider $\bar{F}^{\mu\nu}$ to be real. Hence $\slashed{D}$ turns out to be Hermitian, i.e.\ $\slashed{D}^\dag=\slashed{D}$. 
Now one can follow the method mentioned in \cref{HKE} to obtain the necessary Laplace-type operator $\Lambda$ of the fermionic sector as
{
	\allowdisplaybreaks
	\begin{equation}\label{14}
		\begin{split}
			&\tilde{\xi}_m \Lambda^{mn}\tilde{\xi}_n = \tilde{\xi}_m \slashed{D}^{mp}{\slashed{D}_p}^n\tilde{\xi}_n\\
			&\quad= -\bar{\psi}_{\mu}\Big\lbrace\frac{1}{4}\gamma^\alpha \gamma^\rho\gamma^\mu\gamma^\nu\gamma^\sigma\gamma_\alpha D_\rho D_\sigma- \big(\bar{F}^{\mu\alpha}+\frac{1}{2}\gamma^5 \bar{H}^{\mu\alpha} \big)\big(\bar{F_\alpha}^\nu+\frac{1}{2}\gamma^5{\bar{H_\alpha}}^\nu \big)\Big\rbrace\psi_{\nu}\\
			&\qquad-\bar{\phi}_{\mu}\Big\lbrace\frac{1}{4}\gamma^\alpha \gamma^\rho\gamma^\mu\gamma^\nu\gamma^\sigma\gamma_\alpha D_\rho D_\sigma- \big(\bar{F}^{\mu\alpha}+\frac{1}{2}\gamma^5 \bar{H}^{\mu\alpha} \big)\big(\bar{F_\alpha}^\nu+\frac{1}{2}\gamma^5{\bar{H_\alpha}}^\nu \big)\Big\rbrace\phi_{\nu} \\
			& \qquad-\frac{1}{2}\bar{\psi}_{\mu}\gamma^\alpha\gamma^\rho\gamma^\mu \big(D_\rho\bar{F_\alpha}^\nu+\frac{1}{2}\gamma^5D_\rho{\bar{H_\alpha}}^\nu \big)\phi_{\nu}+\frac{1}{2}\bar{\phi}_{\mu}\gamma^\alpha\gamma^\rho\gamma^\mu \big(D_\rho\bar{F_\alpha}^\nu+\frac{1}{2}\gamma^5D_\rho{\bar{H_\alpha}}^\nu \big)\psi_{\nu}\\
			&\qquad-\frac{1}{2}\bar{\psi}_{\mu}\Big\lbrace \gamma^\alpha\gamma^\rho\gamma^\mu\big(\bar{F_\alpha}^\nu+\frac{1}{2}\gamma^5{\bar{H_\alpha}}^\nu\big) +\big(\bar{F^\mu}_\alpha+\frac{1}{2}\gamma^5{\bar{H^\mu}}_\alpha\big)\gamma^\nu\gamma^\rho\gamma^\alpha\Big\rbrace D_\rho\phi_{\nu}\\
			&\qquad+\frac{1}{2}\bar{\phi}_{\mu}\Big\lbrace \gamma^\alpha\gamma^\rho\gamma^\mu\big(\bar{F_\alpha}^\nu+\frac{1}{2}\gamma^5{\bar{H_\alpha}}^\nu\big) +\big(\bar{F^\mu}_\alpha+\frac{1}{2}\gamma^5{\bar{H^\mu}}_\alpha\big)\gamma^\nu\gamma^\rho\gamma^\alpha\Big\rbrace D_\rho\psi_{\nu}.
		\end{split}
	\end{equation}
}
The second derivative terms of \cref{14} can be simplified to  the  following form \cite{Karan:2018ac},
\begin{align}\label{15}
	&\gamma^\alpha \gamma^\rho\gamma^\mu\gamma^\nu\gamma^\sigma\gamma_\alpha D_\rho D_\sigma = 4\mathbb{I}_4\bar{g}^{\mu\nu}D^\rho D_\rho -(\gamma^\alpha\gamma^\beta\gamma^\mu\gamma^\nu + \gamma^\nu\gamma^\mu\gamma^\beta\gamma^\alpha)[D_\alpha,D_\beta],\nonumber \\
	&\qquad\qquad =4\mathbb{I}_4\bar{g}^{\mu\nu}D^\rho D_\rho-4\mathbb{I}_4R^{\mu\nu}+2\gamma^\alpha\gamma^\beta {R^{\mu\nu}}_{\alpha\beta}-2\gamma^\mu\gamma^\alpha {R^\nu}_\alpha+2\gamma^\nu\gamma^\alpha {R^\mu}_\alpha.
\end{align}
Using \eqref{15} and re-arranging the terms \eqref{14} can be expressed as,
{
	\allowdisplaybreaks
	\begin{align}\label{20}
		\begin{split}
			&\tilde{\xi}_m \Lambda^{mn}\tilde{\xi}_n = -\mathbb{I}_4\bar{g}^{\mu\nu}\bar{\psi}_{\mu}D^\rho D_\rho\psi_{\nu}-\mathbb{I}_4\bar{g}^{\mu\nu}\bar{\phi}_{\mu}D^\rho D_\rho\phi_{\nu}+\bar{\psi}_{\mu}\Big\lbrace\mathbb{I}_4 R^{\mu\nu}-\frac{1}{2}\gamma^\alpha\gamma^\beta {R^{\mu\nu}}_{\alpha\beta}\\
			&\qquad\qquad\quad+\frac{1}{2}\gamma^\mu\gamma^\alpha {R^\nu}_\alpha -\frac{1}{2}\gamma^\nu\gamma^\alpha {R^\mu}_\alpha +\big(\bar{F}^{\mu\alpha}+\frac{1}{2}\gamma^5 \bar{H}^{\mu\alpha}\big)\big(\bar{F_\alpha}^\nu+\frac{1}{2}\gamma^5{\bar{H_\alpha}}^\nu\big)\Big\rbrace\psi_{\nu}\\
			&\qquad\qquad\quad +\bar{\phi}_{\mu}\Big\lbrace\mathbb{I}_4 R^{\mu\nu}-\frac{1}{2}\gamma^\alpha\gamma^\beta {R^{\mu\nu}}_{\alpha\beta}+\frac{1}{2}\gamma^\mu\gamma^\alpha {R^\nu}_\alpha -\frac{1}{2}\gamma^\nu\gamma^\alpha {R^\mu}_\alpha\\
			& \qquad\qquad\quad+\big(\bar{F}^{\mu\alpha}+\frac{1}{2}\gamma^5 \bar{H}^{\mu\alpha}\big)\big(\bar{F_\alpha}^\nu+\frac{1}{2}\gamma^5{\bar{H_\alpha}}^\nu\big)\Big\rbrace\phi_{\nu}\\
			&\qquad\qquad\quad -\frac{1}{2}\bar{\psi}_{\mu}\gamma^\alpha\gamma^\rho\gamma^\mu\big(D_\rho \bar{F_\alpha}^\nu+\frac{1}{2}\gamma^5 D_\rho{\bar{H_\alpha}}^\nu \big)\phi_{\nu}\\
			&\qquad\qquad\quad +\frac{1}{2}\bar{\phi}_{\mu}\gamma^\alpha\gamma^\rho\gamma^\mu\big(D_\rho \bar{F_\alpha}^\nu+\frac{1}{2}\gamma^5 D_\rho{\bar{H_\alpha}}^\nu \big)\psi_{\nu}\\
			&\qquad\qquad\quad -\frac{1}{2}\bar{\psi}_{\mu}\Big\lbrace \gamma^\alpha\gamma^\rho\gamma^\mu\big(\bar{F_\alpha}^\nu+\frac{1}{2}\gamma^5{\bar{H_\alpha}}^\nu\big) +\big(\bar{F^\mu}_\alpha+\frac{1}{2}\gamma^5{\bar{H^\mu}}_\alpha\big)\gamma^\nu\gamma^\rho\gamma^\alpha\Big\rbrace D_\rho\phi_{\nu}\\
			&\qquad\qquad\quad +\frac{1}{2}\bar{\phi}_{\mu}\Big\lbrace \gamma^\alpha\gamma^\rho\gamma^\mu\big(\bar{F_\alpha}^\nu+\frac{1}{2}\gamma^5{\bar{H_\alpha}}^\nu\big) +\big(\bar{F^\mu}_\alpha+\frac{1}{2}\gamma^5{\bar{H^\mu}}_\alpha\big)\gamma^\nu\gamma^\rho\gamma^\alpha\Big\rbrace D_\rho\psi_{\nu}.
		\end{split}
	\end{align}
}
So we have adjusted $\Lambda$ as the prescribed form \eqref{112}, where the matrices $N$ and $P$ hold the following form
{
	\allowdisplaybreaks
	\begin{align}\label{21}
		\begin{split}
			&\tilde{\xi}_m {(N^\rho)}^{mn}\tilde{\xi}_n = \frac{1}{2}\bar{\psi}_{\mu}\Big\lbrace \gamma^\alpha\gamma^\rho\gamma^\mu\big(\bar{F_\alpha}^\nu+\frac{1}{2}\gamma^5{\bar{H_\alpha}}^\nu\big) +\big(\bar{F^\mu}_\alpha+\frac{1}{2}\gamma^5{\bar{H^\mu}}_\alpha\big)\gamma^\nu\gamma^\rho\gamma^\alpha\Big\rbrace \phi_{\nu}\\
			&\qquad\qquad\qquad-\frac{1}{2}\bar{\phi}_{\mu}\Big\lbrace \gamma^\alpha\gamma^\rho\gamma^\mu\big(\bar{F_\alpha}^\nu+\frac{1}{2}\gamma^5{\bar{H_\alpha}}^\nu\big) +\big(\bar{F^\mu}_\alpha+\frac{1}{2}\gamma^5{\bar{H^\mu}}_\alpha\big)\gamma^\nu\gamma^\rho\gamma^\alpha\Big\rbrace\psi_{\nu},\\
			& \tilde{\xi}_m P^{mn}\tilde{\xi}_n = -\bar{\psi}_{\mu}\Big\lbrace\mathbb{I}_4 R^{\mu\nu}-\frac{1}{2}\gamma^\alpha\gamma^\beta {R^{\mu\nu}}_{\alpha\beta}+\frac{1}{2}\gamma^\mu\gamma^\alpha {R^\nu}_\alpha -\frac{1}{2}\gamma^\nu\gamma^\alpha {R^\mu}_\alpha\\
			&\qquad\qquad\qquad +\big(\bar{F}^{\mu\alpha}+\frac{1}{2}\gamma^5 \bar{H}^{\mu\alpha}\big)\big(\bar{F_\alpha}^\nu+\frac{1}{2}\gamma^5{\bar{H_\alpha}}^\nu\big)\Big\rbrace\psi_{\nu}\\
			&\qquad\qquad\qquad -\bar{\phi}_{\mu}\Big\lbrace\mathbb{I}_4 R^{\mu\nu}-\frac{1}{2}\gamma^\alpha\gamma^\beta {R^{\mu\nu}}_{\alpha\beta}+\frac{1}{2}\gamma^\mu\gamma^\alpha {R^\nu}_\alpha -\frac{1}{2}\gamma^\nu\gamma^\alpha {R^\mu}_\alpha\\
			&\qquad\qquad\qquad +\big(\bar{F}^{\mu\alpha}+\frac{1}{2}\gamma^5 \bar{H}^{\mu\alpha}\big)\big(\bar{F_\alpha}^\nu+\frac{1}{2}\gamma^5{\bar{H_\alpha}}^\nu\big)\Big\rbrace\phi_{\nu}\\
			&\qquad\qquad\qquad +\frac{1}{2}\bar{\psi}_{\mu}\gamma^\alpha\gamma^\rho\gamma^\mu\big(D_\rho \bar{F_\alpha}^\nu+\frac{1}{2}\gamma^5 D_\rho{\bar{H_\alpha}}^\nu \big)\phi_{\nu}\\
			&\qquad\qquad\qquad-\frac{1}{2}\bar{\phi}_{\mu}\gamma^\alpha\gamma^\rho\gamma^\mu\big(D_\rho \bar{F_\alpha}^\nu+\frac{1}{2}\gamma^5 D_\rho{\bar{H_\alpha}}^\nu \big)\psi_{\nu}.
		\end{split}
	\end{align} 
}
As mentioned in \cref{HKE}, now our task is to express $\Lambda$ as the standard form \eqref{113}. Then, using the formulae \eqref{114} we find
{
	\allowdisplaybreaks
	\begin{align}\label{23}
		\begin{split}
			&\tilde{\xi}_m {(\omega^\rho)}^{mn}\tilde{\xi}_n=  \frac{1}{4}\bar{\psi}_{\mu}\Big\lbrace \gamma^\alpha\gamma^\rho\gamma^\mu\big(\bar{F_\alpha}^\nu+\frac{1}{2}\gamma^5{\bar{H_\alpha}}^\nu\big) +\big(\bar{F^\mu}_\alpha+\frac{1}{2}\gamma^5{\bar{H^\mu}}_\alpha\big)\gamma^\nu\gamma^\rho\gamma^\alpha\Big\rbrace \phi_{\nu}\\
			&\qquad\qquad\qquad-\frac{1}{4}\bar{\phi}_{\mu}\Big\lbrace \gamma^\alpha\gamma^\rho\gamma^\mu\big(\bar{F_\alpha}^\nu+\frac{1}{2}\gamma^5{\bar{H_\alpha}}^\nu\big) +\big(\bar{F^\mu}_\alpha+\frac{1}{2}\gamma^5{\bar{H^\mu}}_\alpha\big)\gamma^\nu\gamma^\rho\gamma^\alpha\Big\rbrace\psi_{\nu},
		\end{split}
	\end{align}
}
and also identify the corresponding  $I$, $E$ and $\Omega$ as the following compact form 
{
	\allowdisplaybreaks
	\begin{align}\label{24}
		\begin{split}
			&\tilde{\xi}_m I^{mn}\tilde{\xi}_n = \bar{\psi}_{\mu}\mathbb{I}_4 \bar{g}^{\mu\nu}\psi_{\nu}+\bar{\phi}_{\mu}\mathbb{I}_4 \bar{g}^{\mu\nu}\phi_{\nu},\\
			&\tilde{\xi}_m E^{mn}\tilde{\xi}_n= \tilde{\xi}_m P^{mn}\tilde{\xi}_n -\tilde{\xi}_m(D^\rho\omega_\rho)^{mn}\tilde{\xi}_n-\tilde{\xi}_m(\omega^\rho)^{mp}{(\omega_\rho)_p}^n\tilde{\xi}_n,\\
			&\tilde{\xi}_m (\Omega_{\rho\sigma})^{mn}\tilde{\xi}_n = \bar{\psi}_{\mu}\big(\mathbb{I}_4{R^{\mu\nu}}_{\rho\sigma}+\frac{1}{4}\bar{g}^{\mu\nu}\gamma^\alpha\gamma^\beta R_{\rho\sigma\alpha\beta}\big)\psi_{\nu}\\
			&\qquad\qquad\qquad\quad +\bar{\phi}_{\mu}\big(\mathbb{I}_4{R^{\mu\nu}}_{\rho\sigma}+\frac{1}{4}\bar{g}^{\mu\nu}\gamma^\alpha\gamma^\beta R_{\rho\sigma\alpha\beta}\big)\phi_{\nu}\\
			&\qquad\qquad\qquad\quad+ \tilde{\xi}_m{D_{[\rho}\omega_{\sigma]}}^{mn}\tilde{\xi}_n+\tilde{\xi}_m[\omega_\rho,\omega_\sigma]^{mn}\tilde{\xi}_n.
		\end{split}
	\end{align}
}
We then computed all the necessary trace values using the information \eqref{24} to obtain the Seeley-DeWitt coefficients for the gravitini fields,
\begin{align}\label{gravitini}
	\begin{split}
		&\text{tr}(I) =16+16= 32,\\
		&\text{tr}(E) = -16\bar{F}_{\mu\nu}\bar{F}^{\mu\nu},\\
		&\text{tr}(E^2) = 4R_{\mu\nu\rho\sigma}R^{\mu\nu\rho\sigma}+32R_{\mu\nu}R^{\mu\nu}-8R_{\mu\nu\rho\sigma}\bar{F}^{\mu\nu}\bar{F}^{\rho\sigma}+16(\bar{F}_{\mu\nu}\bar{F}^{\mu\nu})^2,\\
		&\text{tr}(\Omega_{\rho\sigma}\Omega^{\rho\sigma})  = -12R_{\mu\nu\rho\sigma}R^{\mu\nu\rho\sigma}-144R_{\mu\nu}R^{\mu\nu}+48R_{\mu\nu\rho\sigma}\bar{F}^{\mu\nu}\bar{F}^{\rho\sigma} -96(\bar{F}_{\mu\nu}\bar{F}^{\mu\nu})^2.
	\end{split}
\end{align}
Although these computations are extremely lengthy, one can proceed in a systematic way (refer \Cref{A1} for details of this computation).
Making use of these trace results \eqref{gravitini}, one  computes the first three Seeley-DeWitt coefficients \eqref{115} for the fermions (two gravitini) in our $\mathcal{N}=2$ SUGRA. The results are
\begin{align}\label{29}
	\begin{split}
		&(4\pi)^2 a_0^{\text{gravitini}} = -16,\\
		&(4\pi)^2 a_2^{\text{gravitini}} = 8 \bar{F}_{\mu\nu}\bar{F}^{\mu\nu},\\
		&(4\pi)^2 a_4^{\text{gravitini}} =-\frac{1}{90}\left(53 R_{\mu\nu\rho\sigma}R^{\mu\nu\rho\sigma}+172 R_{\mu\nu}R^{\mu\nu} \right).
	\end{split}
\end{align}
Note that we have put an overall $-1/2$ factor by hand on each coefficient to justify the fermion and Majorana condition, as discussed in \cref{HKE}.

Next, the choice of gauge fixing in \cref{11} introduces six bosonic ghosts in our system with the action,
\begin{equation}
	\mathcal{L}_{\text{ghost},f}= \bar{\tilde{b}}_A\gamma^\mu D_\mu \tilde{c}_A + \bar{\tilde{e}}_A\gamma^\mu D_\mu \tilde{e}_A,
\end{equation}
where $\tilde{b}_A$, $\tilde{c}_A$, and $\tilde{e}_A$ are spin half ghosts with $A=1,2$ for the two gravitini species.
These Majorana ghosts are minimally coupled, hence one can obtain the Seeley-DeWitt coefficients as $6$ times the spin-$\frac{1}{2}$ Seeley-DeWitt coefficients with an opposite sign, i.e.\ ${a_{2n}}^{ghost,f}=-6a_{2n}^{1/2}$. The detailed calculations for $a_{2n}^{1/2}$ have been described in \cite{Karan:2018ac}\footnote{See § 2.2.1 of \cite{Karan:2018ac} and finally, use its eq. 2.19 with $R=0$ for $a_{2n}^{1/2}$.} for a general background, and it is very much straightforward. This led us to express the results for ghost part of fermionic fields as 
\begin{align}\label{30}
	\begin{split}
		&(4\pi)^2 a_0^{\text{ghost},f} = 12,\\
		&(4\pi)^2 a_2^{\text{ghost},f} = 0,\\
		&(4\pi)^2 a_4^{\text{ghost},f} =-\frac{1}{120}\left(7 R_{\mu\nu\rho\sigma}R^{\mu\nu\rho\sigma}+8 R_{\mu\nu}R^{\mu\nu} \right).
	\end{split}
\end{align}  
\subsection{Combined Seeley-DeWitt coefficients in minimal $\mathcal{N}=2$ EMSGT}
In previous subsections, we have calculated Seeley-DeWitt coefficients for the field fluctuations of the bosonic and fermionic part of $\mathcal{N}=2$ SUGRA and also computed the same for corresponding ghost parts separately. These ghost fields have no interaction with any of graviton, graviphoton and gravitini fields. Therefore, the contributions from the ghost fields can be directly added to the results from the  gauge fixed part. The total Seeley-DeWitt coefficients are obtained by adding the bosonic and fermionic contributions.

Combining the results \eqref{43} and \eqref{50} we get the Seeley-DeWitt coefficients for the bosonic sector of $\mathcal{N}=2$ SUGRA as
\begin{align}\label{51}
	\begin{split}
		&(4\pi)^2 a_0^{b} = 4,\\
		&(4\pi)^2 a_2^{b} = 6 \bar{F}_{\mu\nu}\bar{F}^{\mu\nu},\\
		&(4\pi)^2 a_4^{b} = \frac{1}{180}\left(199 R_{\mu\nu\rho\sigma}R^{\mu\nu\rho\sigma}+26 R_{\mu\nu}R^{\mu\nu} \right) .
	\end{split}
\end{align}
This clearly reflects the result derived in \cite{Bhattacharyya:2012ss}.
Similarly, for the fermionic part, we sum up the results \eqref{29} and \eqref{30}. This yields
\begin{align}\label{31}
	\begin{split}
		&(4\pi)^2 a_0^{f} = -4,\\
		&(4\pi)^2 a_2^{f} = 8 \bar{F}_{\mu\nu}\bar{F}^{\mu\nu},\\
		&(4\pi)^2 a_4^{f} = -\frac{1}{360}\left(233 R_{\mu\nu\rho\sigma}R^{\mu\nu\rho\sigma}+712 R_{\mu\nu}R^{\mu\nu} \right) .
	\end{split}
\end{align}
Our results for the fermionic sector also agrees with eigenfunction approach results \cite{Sen:2012qq} for $\text{AdS}_2\times S^2$ background, as well as with the hybrid approach results \cite{Charles:2015nn}.\footnote{Use the identities (4.2) in the results (3.84) of \cite{Charles:2015nn}.} 
We then find the full $\mathcal{N}=2$ SUGRA Seeley-DeWitt coefficients with the contribution graviton, graviphoton, and gravitini fluctuations as $a^{\mathcal{N}=2}_{2n}=a_{2n}^{b}+a_{2n}^{f}$. The results are
\begin{align}\label{52}
	\begin{split}
		&(4\pi)^2 a^{\mathcal{N}=2}_0 = 0,\\
		&(4\pi)^2 a^{\mathcal{N}=2}_2 = 14 \bar{F}_{\mu\nu}\bar{F}^{\mu\nu},\\
		&(4\pi)^2 a^{\mathcal{N}=2}_4 = \frac{1}{24}\left(11 R_{\mu\nu\rho\sigma}R^{\mu\nu\rho\sigma}-44 R_{\mu\nu}R^{\mu\nu} \right) .
	\end{split}
\end{align}
The following section deals with
the use of these coefficients in the logarithmic correction to the entropy of extremal black holes.
\section{Application of Seeley-DeWitt coefficients: logarithmic corrections to the black hole entropy}\label{app}
In this section, we will summarize the general procedure to find the logarithmic correction to extremal back holes in an arbitrary theory by making use of the Seeley-DeWitt coefficient $a_4(x)$ of the same theory \cite{Sen:2008wa,Sen:2009wb,Sen:2009wc,Gibbons:1977ta}.
Then we will use this procedure for extremal black holes in minimal $\mathcal{N}$=2 Einstein-Maxwell supergravity theory.
It has been observed that Euclidean quantum gravity \cite{Gibbons:1977ta} gives a more detailed picture of the quantum corrections \cite{Banerjee:2011oo,Banerjee:2011pp,Sen:2012rr} to the black hole entropy. We will follow the quantum entropy function formalism \cite{Sen:2008wa,Sen:2009wb,Sen:2009wc},
in order to calculate the logarithmic corrections to the entropy of string theoretic black holes. 
For extremal black holes, $\text{AdS}_2/\text{CFT}_1$ correspondence yields an auxiliary way of describing the horizon degeneracy: they are equivalently given by finite (boundary independent) part of the partition function $Z_{\text{AdS}_2 \times \mathcal{K}}$  of the string theory in the near horizon geometry structured as $\text{AdS}_2 \times \mathcal{K}$, where $\mathcal{K}$ is compact space fibered over $\text{AdS}_2$. The classical limit provides us with the relation: $Z_{\text{AdS}_2 \times \mathcal{K}}^{\text{finite}}=e^{S_{\text{BH}}}$, where $S_{\text{BH}}$ is the quantum corrected black hole entropy. The partition function $Z_{\text{AdS}_2 \times \mathcal{K}}$ computed by carrying out Euclidean path integral in $\text{AdS}_2 \times \mathcal{K}$ weighted by $e^{-\mathcal{S}}$ ($\mathcal{S}$ is Euclidean action), over all string fields that asymptotically approaches to the near horizon geometry of black hole and for $Z_{\text{AdS}_2 \times \mathcal{K}}^{\text{finite}}$ one needs to consider only one-loop term which is finite and boundary independent \cite{Banerjee:2011oo,Banerjee:2011pp,Sen:2012rr,Sen:2012qq}. 
Logarithmic corrections ($\Delta S_{\text{BH}}$) are special type of one-loop quantum corrections, requires integrating out non-zero modes of massless fields\footnote{In this context, we will only turn on massless fields like two-dimensional metric $g_{\mu\nu}$, U(1) gauge fields $A_\mu$, etc. among other fields present in $\text{AdS}_2 \times \mathcal{K}$.} in $Z_{\text{AdS}_2 \times \mathcal{K}}$, and then find one-loop effective action $W$ or effective Lagrangian density $\Delta\mathcal{L}_{\text{eff}}$ in quadratic order fluctuation of $\mathcal{S}$ (see \cref{heat kernel}) in the background $\text{AdS}_2 \times \mathcal{K}$. We find logarithmic corrections as
\begin{equation}\label{54}
	\Delta S_{\text{BH}} = -W= -\int_{\text{AdS}_2 \times \mathcal{K}} d^4x \sqrt{\text{det}\thinspace \bar{g}}\thinspace \Delta\mathcal{L}_{\text{eff}}^{(\text{nz})}.
\end{equation}
Here $\mathcal{L}_{\text{eff}}^{(\text{nz})}$ is the one-loop corrected effective Lagrangian density with the contribution of only non-zero modes of massless fields, i.e.\ one needs to throw away zero-modes contribution from the overall result. Then we express \cite{Sen:2012qq,Sen:2012rr}
\begin{align}
	\Delta\mathcal{L}_{\text{eff}}^{(\text{nz})} &=-\frac{\chi}{2}\int_\epsilon^\infty \frac{ds}{s}\Big( K(x,x;s)-\bar{K}(x,x;0)\Big)\\
	&\simeq -\frac{1}{2}\Big(a_4(x)- \chi \bar{K}(x,x;0)\Big)\text{ln}\thinspace A_{H}, \label{55}
\end{align}
where $\bar{K}(x,x;0)$ is the zero mode contribution in $ K(x,x;s)$.\footnote{We denote zero mode ($\lambda_i = 0$) eigenfunctions as $\lbrace g_i^m \rbrace$. Then following \eqref{208}, we define $\bar{K}^{mn}(x,x;0) = \sum_i g_i^m(x)g_i^n(x),\enspace \bar{K}(x,x;0)=G_{mn}\bar{K}^{mn}(x,x;0) $, with the same orthonormal condition as \eqref{209}. The number of zero modes is defined as $n_{\text{zm}}=-\int d^4x \sqrt{\text{det}\thinspace \bar{g}}\thinspace \chi \bar{K}(x,x;0)$.} Note that logarithmic part arises only from the $s$ independent part of the expansion \eqref{108} of $K(x,x;s)$ after integrating over the range $\epsilon\ll s\ll A_H$, where $\epsilon\sim G_N=1/{16\pi}$. Finally, we express the key formula for finding  logarithmic correction to extremal black hole entropy as
\begin{equation}\label{56}
	\Delta S_{\text{BH}} = \frac{1}{2}(C_{\text{local}}+C_{\text{zm}})\text{ln}\thinspace A_{H},
\end{equation}
where the non-zero mode and zero mode contributions are respectively,
\begin{align}
	&C_{\text{local}} = \int_{\text{AdS}_2 \times \mathcal{K}} d^4x \sqrt{\text{det}\thinspace \bar{g}}\thinspace a_4(x),\label{58}\\
	& C_{\text{zm}}= \int_{\text{AdS}_2 \times \mathcal{K}} d^4x \sqrt{\text{det}\thinspace \bar{g}}\thinspace \chi (\beta-1)\bar{K}(x,x;0),\label{59}
\end{align}
where $\beta$ is the scaling dimension of fields arises for incorporating corrections due to zero mode contribution in $\Delta S_{\text{BH}}$ \cite{Sen:2012qq,Sen:2012rr}. One can evaluate $C_{\text{local}}$ using by computing $a_4$ of massless fields in near horizon geometry of concerned black hole, while $C_{\text{zm}}$ can be obtained from the following special combined formula \cite{Charles:2015nn}, 
\begin{equation}\label{60}
	C_{\text{zm}} = -(3+\mathbb{K}) + 2N_{\text{SUSY}}.
\end{equation}
Here $\mathbb{K}$ (number of rotational isometries) is 3 if for black holes having $J=0$, otherwise it is 1; $N_{\text{SUSY}}$ (number of preserved supercharges) is 4 for BPS black holes, 0 otherwise. In order to compute the logarithmic correction to extremal black hole entropy in any arbitrary theory, one needs to first compute $a_4(x)$ for the quadratic fluctuations of massless fields in the theory. Then find the results for $C_{\text{local}}$ (using the $a_4(x)$ result) and $C_{\text{zm}}$ respectively, and finally use these results in the formula \eqref{56}. With cognizance of the above structure, our next aim to compute logarithmic corrections to the entropy of extremal Kerr-Newman, Reissner-Nordstr\"om and Kerr black holes\footnote{Only extremal Reissner-Nordstr\"om black holes are half-BPS in the four-dimensional $\mathcal{N}=2$ EMSGT.} in minimal $\mathcal{N}=2$ EMSGT by making use of $a_4(x)$ result \eqref{52} of the theory. 

\subsection{Logarithmic corrections to $\mathcal{N}=2$ extremal black hole entropy}\label{lc}
We will start with a Kerr-Newman black hole \cite{Bhattacharyya:2012ss} to calculate associated logarithmic correction.
In appropriate units, the metric of a general Kerr-Newman black hole characterized by mass $M$, charge $Q$, and angular momentum $J$ is expressed as 
{
	\allowdisplaybreaks
	\begin{equation}\label{61}
		\begin{split}
			ds^2 &= -\frac{r^2+a^2\text{cos}^2\theta-2Mr+Q^2}{r^2+a^2\text{cos}^2\theta}dt^2+\frac{r^2+a^2\text{cos}^2\theta}{r^2+a^2-2Mr+Q^2}dr^2 + (r^2+a^2\text{cos}^2\theta)d\theta^2\\
			&\enspace +\frac{(r^2+a^2\text{cos}^2\theta)(r^2+a^2)+(2Mr-Q^2)a^2\text{sin}^2\theta}{r^2+a^2\text{cos}^2\theta}\text{sin}^2\theta d\varphi^2\\
			&\enspace +\frac{2(Q^2-2Mr)a}{r^2+a^2\text{cos}^2\theta}\text{sin}^2\theta dt d\varphi,
		\end{split}
\end{equation}}
where $a=J/M$. The horizon is situated at 
\begin{equation}\label{62}
	r= r_H  = M+ \sqrt{M^2-Q^2-a^2},
\end{equation}
and the corresponding classical Bekenstein-Hawking entropy of black hole having a horizon area $A_H = 4\pi({r_H}^2+a^2)$ is\footnote{The classical black hole entropy formula is given as $S_{\text{BH}} = \frac{A_H}{4G_N}$, where $G_N = 1/16\pi$ in our convention.}
\begin{equation}\label{63}
	S_{\text{BH}} = 16\pi^2\left(2M^2-Q^2+2M\sqrt{M^2-a^2-Q^2} \right).
\end{equation}
The extremal limit $M\to \sqrt{a^2+Q^2}$ yields
\begin{equation}\label{64}
	S_{\text{BH}} = 16\pi^2(2a^2+Q^2).
\end{equation}
Now for the metric \eqref{61}, we have \cite{Henry:2000wd,Cherubini:2002we}
{
	\allowdisplaybreaks
	\begin{align}\label{65}
		\begin{split}
			R_{\mu\nu\rho\sigma}R^{\mu\nu\rho\sigma} &= \frac{8}{(r^2+a^2\text{cos}^2\theta)^6}\Big\lbrace 6M^2(r^6-15a^2r^4\text{cos}^2\theta+ 15a^4r^2\text{cos}^4\theta\\
			&\enspace -a^6\text{cos}^6\theta)-12MQ^2r(r^4-10r^2a^2\text{cos}^2\theta+ 5a^4\text{cos}^4\theta)\\
			&\enspace  +Q^4(7r^4-34r^2a^2\text{cos}^2\theta+7a^4\text{cos}^4\theta) \Big\rbrace,\\
			R_{\mu\nu}R^{\mu\nu}&= \frac{4Q^4}{(r^2+a^2\text{cos}^2\theta)^4},\\
			\text{det}\thinspace \bar{g} &= (r^2+a^2\text{cos}^2\theta)^2\text{sin}^2\theta.
		\end{split}
	\end{align}
}
In the extremal limit, one can achieve \cite{Bhattacharyya:2012ss} 
{
	\allowdisplaybreaks
	\begin{align}\label{66}
		\begin{split}
			\int_{\text{horizon}} d^4x \sqrt{\text{det}\thinspace \bar{g}}\thinspace R_{\mu\nu\rho\sigma}R^{\mu\nu\rho\sigma} &= -\frac{16\pi^2}{b(b^2+1)^{5/2}(2b^2+1)}\Big\lbrace 3(2b^2+1)^2 \text{tan}^{-1}\left(\frac{b}{\sqrt{b^2+1}} \right)\\
			&\quad +b\sqrt{b^2+1}(-8b^6-20b^4-8b^2+1)\Big\rbrace,\\
			\int_{\text{horizon}} d^4x \sqrt{\text{det}\thinspace \bar{g}}R_{\mu\nu}R^{\mu\nu}&= -\frac{4\pi^2}{b(b^2+1)^{5/2}(2b^2+1)}\Big\lbrace 3(2b^2+1)^2 \text{tan}^{-1}\left(\frac{b}{\sqrt{b^2+1}} \right)\\
			&\quad +b\sqrt{b^2+1}(8b^2+5)\Big\rbrace,
		\end{split}
	\end{align}
}
where $b=a/Q= J/MQ$.
\begin{itemize}
	\item[\scalebox{0.6}{$\blacksquare$}]
	Using the $a_4(x)$ result \eqref{52} for minimal $\mathcal{N}=2$ EMSGT and the limits \eqref{66} in the relation \eqref{58}, and also making use of the formula \eqref{60}, for extremal \textit{Kerr-Newman} black hole ($J \neq 0, Q\neq 0$) we have
	\begin{equation}\label{67}
		C_{\text{local}} = \frac{11}{6}, \enspace C_{\text{zm}} = -(3+1)+2\times 0=-4.
	\end{equation}
	Substitution of these in the formula \eqref{56} yields
	\begin{equation}\label{68}
		\Delta S_{\text{BH}} = -\frac{13}{12}\thinspace \text{ln}\thinspace A_{H}.
	\end{equation}
	This is one of our central results. Similarly, we can get the results for Kerr and Reissner-Nordstr\"om black holes by applying appropriate limits as discussed below.
	\item[\scalebox{0.6}{$\blacksquare$}]
	For an extremal \textit{Kerr} black hole ($J\neq 0, Q=0$), we have to take $b\to\infty$ in the limits of background invariants given in \eqref{66}. This yields
	\begin{equation}\label{69}
		C_{\text{local}}= \frac{11}{6}, \enspace C_{\text{zm}} = -(3+1)+2\times 0=-4,
	\end{equation}
	hence the logarithmic correction becomes
	\begin{equation}\label{70}
		\Delta S_{\text{BH}} = -\frac{13}{12}\thinspace \text{ln}\thinspace A_{H}.
	\end{equation}
	This is in accordance with the result in \cite{Sen:2012rr} (set $n_V =1, n_{3/2}=2, n_S= n_F = 0$ in eq.(3.12) and include an additional $-4\thinspace \text{ln}\thinspace A_{H}$ term for the correction due to zero-modes of gravitini).
	\item[\scalebox{0.6}{$\blacksquare$}]
	For the case of an extremal \textit{Reissner-Nordstr\"om} black hole ($J=0, Q\neq 0$), one has to put the limit $b\to 0$ in the limits of background invariants given in \eqref{66}. This gives
	\begin{equation}\label{71}
		C_{\text{local}}= \frac{11}{6}, \enspace C_{\text{zm}} = -(3+3)+2\times 4=2,
	\end{equation}
	hence the logarithmic correction becomes
	\begin{equation}\label{72}
		\Delta S_{\text{BH}} = \frac{23}{12}\thinspace \text{ln}\thinspace A_{H}.
	\end{equation}
	This agrees with the result in \cite{Sen:2012qq} (see eq.(5.32)).
\end{itemize}

To conclude, we have computed the first three Seeley-DeWitt coefficients for minimal $\mathcal{N}=2$ Einstein-Maxwell supergravity theory following the general Seeley-DeWitt approach \cite{Vassilevich:2003ll} (reviewed in \cref{HKE}). The fermionic sector results are newly calculated in this approach, while the bosonic sector results are reviewed and recalculated from \cite{Bhattacharyya:2012ss}. This computation does not impose any limitations on the background geometry. The coefficients are expressed in terms of the local invariants of the theory and can be evaluated for any arbitrary background. These Seeley-DeWitt coefficient results for the minimal $\mathcal{N}=2$ Einstein-Maxwell supergravity theory in four dimensions agree with eigenfunction approach results \cite{Sen:2012qq} for $\text{AdS}_2\times S^2$ background, as well as with the hybrid approach results \cite{Charles:2015nn}. As an application of these results, we have computed the logarithmic corrections to Bekenstein-Hawking entropy of the extremal Kerr-Newman, Kerr and Reissner-Nordstr\"om black holes in minimal $\mathcal{N}=2$ Einstein-Maxwell supergravity theory by properly integrating these coefficients in the near horizon geometry of these black holes. The Kerr and Reissner-Nordstr\"om results agree with \cite{Sen:2012rr} and \cite{Sen:2012qq} respectively, while the Kerr-Newman result is new to the literature. Any microscopic theory explaining the entropy of such black holes is expected to reproduce the subleading logarithmic correction terms specified in \cref{68} in addition to the leading Bekenstein-Hawking term.

\acknowledgments

It is a great pleasure to thank Ashoke Sen for useful discussions at different stages of this work. The authors would like to thank Finn Larsen for sharing insight and expertise that greatly improved this manuscript. We would also like to acknowledge NISER, HRI, BHU and IISER-TVM for their hospitality. BP acknowledges IIT(ISM), Dhanbad for the Grant (FRS (53)/2013-2014/APH). 

\appendix

\section{Trace calculations}\label{A1}
Using Equations of motion and identities as mentioned in \cref{EOM}, we derive the following identities\footnote{We have assumed the Euclidean signature for all the derivations in this paper.}
{
	\allowdisplaybreaks
	\begin{align}\label{id1}
		\begin{split}
			&\bar{H}_{\mu\nu}\bar{H}^{\mu\nu} = 4 \bar{F}_{\mu\nu}\bar{F}^{\mu\nu}, \thinspace \varepsilon^{\mu\nu\rho\sigma}\bar{H}_{\rho\sigma}=-4\bar{F}^{\mu\nu},\\
			&\bar{H}_{\mu\rho}{\bar{H_\nu}}^\rho = -2R_{\mu\nu}+\bar{g}_{\mu\nu}\bar{F}_{\rho\sigma}\bar{F}^{\rho\sigma},\\
			&\bar{F}^{\mu\rho}\bar{F^\nu}_\rho \bar{H}_{\mu\sigma}{\bar{H_\nu}}^\sigma = \bar{F}^{\mu\rho}{\bar{H^\nu}}_\rho \bar{H}_{\mu\sigma}\bar{F_\nu}^\sigma= (\bar{F}_{\mu\nu}\bar{F}^{\mu\nu})^2-R_{\mu\nu}R^{\mu\nu},\\
			&\bar{F}_{\mu\nu}\bar{F}_{\rho\sigma}\bar{H}^{\mu\nu}\bar{H}^{\rho\sigma}= 4\left((\bar{F}_{\mu\nu}\bar{F}^{\mu\nu})^2-R_{\mu\nu}R^{\mu\nu}\right),\\
			&R_{\mu\rho\nu\sigma}\bar{F}^{\mu\nu}\bar{F}^{\rho\sigma}= \frac{1}{2}R_{\mu\nu\rho\sigma}\bar{F}^{\mu\nu}\bar{F}^{\rho\sigma},\\
			&R_{\mu\rho\nu\sigma}\bar{H}^{\mu\nu}\bar{H}^{\rho\sigma}= 2\left(R_{\mu\nu\rho\sigma}\bar{F}^{\mu\nu}\bar{F}^{\rho\sigma}-2R_{\mu\nu}R^{\mu\nu} \right),\\
			&R_{\mu\nu\rho\sigma}\bar{H}^{\mu\nu}\bar{H}^{\rho\sigma}= 4\left(R_{\mu\nu\rho\sigma}\bar{F}^{\mu\nu}\bar{F}^{\rho\sigma}-2R_{\mu\nu}R^{\mu\nu} \right),
		\end{split}
	\end{align}
}
followed by
{
	\allowdisplaybreaks
	\begin{align}\label{id2}
		\begin{split}
			&(D_\rho \bar{F}_{\mu\nu})(D^\rho \bar{F}^{\mu\nu}) = R_{\mu\nu\rho\sigma}\bar{F}^{\mu\nu}\bar{F}^{\rho\sigma}-R_{\mu\nu}R^{\mu\nu},\\
			&(D_\rho \bar{H}_{\mu\nu})(D^\rho \bar{H}^{\mu\nu}) = 4\left(R_{\mu\nu\rho\sigma}\bar{F}^{\mu\nu}\bar{F}^{\rho\sigma}-R_{\mu\nu}R^{\mu\nu}\right),\\
			&(D_\mu \bar{{F}_\rho}^\nu)(D_\nu \bar{F}^{\rho\mu}) = \frac{1}{2}\left(R_{\mu\nu\rho\sigma}\bar{F}^{\mu\nu}\bar{F}^{\rho\sigma}-R_{\mu\nu}R^{\mu\nu}\right),\\
			&(D_\mu {\bar{H_\rho}}^\nu)(D_\nu \bar{H}^{\rho\mu}) = 2\left(R_{\mu\nu\rho\sigma}\bar{F}^{\mu\nu}\bar{F}^{\rho\sigma}-R_{\mu\nu}R^{\mu\nu}\right).
		\end{split}
	\end{align}
}
We note that in our analysis we will ignore total derivatives terms as these will not contribute in heat kernel coefficient integrals. In deriving \cref{id2} we will apply the same, also use Maxwell equation and perform the following steps
\begin{equation}
	\begin{split}
		(D_\rho \bar{F}_{\mu\nu})(D^\rho \bar{F}^{\mu\nu}) &= 2\bar{F^\mu}_\nu D_\rho D_\mu \bar{F}^{\nu\rho},\\
		&= 2\bar{F^\mu}_\nu [D_\rho,D_\mu] \bar{F}^{\nu\rho}.
	\end{split}
\end{equation}
We will do the same for dual field strength $\bar{H}_{\mu\nu}$. Finally, the covariant derivative commutator acting on rank-2 tensor gives the form of the result \eqref{id2}.

Also, using \cref{23} for the two gravitini species we write,\footnote{Here only $\mu,\nu$ are particular tensor indices, labelled by the combination of $\psi$ and $\phi$ fields to denote with which set of field combination is the particular $\omega$ associated with (or $\mu,\nu$ are contracted with). Similar notation will be used for $E$ and $\Omega$ also.}
\begin{subequations}
	\begin{align}
		(\omega^\rho)^{\psi_\mu\phi_\nu} &= \frac{1}{4}\gamma^\alpha\gamma^\rho\gamma^\mu\Big(\bar{F_\alpha}^\nu+\frac{1}{2}\gamma^5{\bar{H_\alpha}}^\nu\Big) + \frac{1}{4}\Big(\bar{F^\mu}_\alpha+\frac{1}{2}\gamma^5 \bar{H^\mu}_\alpha\Big)\gamma^\nu\gamma^\rho\gamma^\alpha \label{a1},\\
		(\omega^\rho)^{\phi_\mu\psi_\nu} &=-\frac{1}{4}\gamma^\alpha\gamma^\rho\gamma^\mu\Big(\bar{F_\alpha}^\nu+\frac{1}{2}\gamma^5{\bar{H_\alpha}}^\nu\Big) - \frac{1}{4}\Big(\bar{F^\mu}_\alpha+\frac{1}{2}\gamma^5 \bar{H^\mu}_\alpha\Big)\gamma^\nu\gamma^\rho\gamma^\alpha.\label{a2}
	\end{align}
\end{subequations}
\subsubsection*{ Calculating $\text{tr}(E)$ and $\text{tr}(E^2)$}
\begin{align}
	\text{tr}(E) &= {E^{\psi_\mu}}_{\psi_\mu}+ {E^{\phi_\mu}}_{\phi_\mu}+ {E^{\psi_\mu}}_{\phi_\mu}+ {E^{\phi_\mu}}_{\psi_\mu},\\
	\text{tr}(E^2) &= {E^{\psi_\mu}}_{\psi_\nu}{E^{\psi_\nu}}_{\psi_\mu}+ {E^{\phi_\mu}}_{\phi_\nu}{E^{\phi_\nu}}_{\phi_\mu}+ {E^{\psi_\mu}}_{\phi_\nu}{E^{\phi_\nu}}_{\psi_\mu}+ {E^{\phi_\mu}}_{\psi_\nu}{E^{\psi_\nu}}_{\phi_\mu}.
\end{align}
Making use of expressions of $\omega$ and $P$ from \cref{23} and \cref{21} respectively, one can simplify $E$ to the following form
{
	\allowdisplaybreaks
	\begin{align}\label{25}
		\begin{split}
			&\tilde{\xi}_m E^{mn}\tilde{\xi}_n= \bar{\psi}_{\mu}{E^{\psi_\mu \psi_\nu}}\psi_{\nu}+ \bar{\phi}_{\mu}{E^{\phi_\mu \phi_\nu}}\phi_{\nu}+\bar{\psi}_{\mu}{E^{\psi_\mu \phi_\nu}}\phi_{\nu}+ \bar{\phi}_{\mu}{E^{\phi_\mu \psi_\nu}}\psi_{\nu},\\
			&\qquad\qquad =\bar{\psi}_{\mu}\Big\lbrace- \mathbb{I}_4 R^{\mu\nu}+\frac{1}{2}\gamma^\alpha\gamma^\beta {R^{\mu\nu}}_{\alpha\beta}-\frac{1}{2}\gamma^\mu\gamma^\alpha {R^\nu}_\alpha +\frac{1}{2}\gamma^\nu\gamma^\alpha {R^\mu}_\alpha\\
			&\quad\qquad\qquad +\frac{1}{2}\gamma^\alpha\gamma^\beta\big(\bar{F}_{\alpha\beta}-\frac{1}{2}\gamma^5\bar{H}_{\alpha\beta}\big)\big(\bar{F}^{\mu\nu}+\frac{1}{2}\gamma^5 \bar{H}^{\mu\nu}\big)\\
			&\quad\qquad\qquad +\frac{1}{4}\bar{g}^{\mu\nu}\gamma^\alpha\gamma^\beta\big(\bar{F_\alpha}^\tau-\frac{1}{2}\gamma^5 \bar{H_\alpha}^\tau\big)\big(\bar{F}_{\tau\beta}-\frac{1}{2}\gamma^5 \bar{H}_{\tau\beta}\big)\Big\rbrace \psi_{\nu}\\
			&\quad\qquad\qquad +\bar{\phi}_{\mu}\Big\lbrace- \mathbb{I}_4 R^{\mu\nu}+\frac{1}{2}\gamma^\alpha\gamma^\beta {R^{\mu\nu}}_{\alpha\beta}-\frac{1}{2}\gamma^\mu\gamma^\alpha {R^\nu}_\alpha +\frac{1}{2}\gamma^\nu\gamma^\alpha {R^\mu}_\alpha\\
			&\quad\qquad\qquad +\frac{1}{2}\gamma^\alpha\gamma^\beta\big(\bar{F}_{\alpha\beta}-\frac{1}{2}\gamma^5\bar{H}_{\alpha\beta}\big)\big(\bar{F}^{\mu\nu}+\frac{1}{2}\gamma^5 \bar{H}^{\mu\nu}\big)\\
			&\quad\qquad\qquad +\frac{1}{4}\bar{g}^{\mu\nu}\gamma^\alpha\gamma^\beta\big(\bar{F_\alpha}^\tau-\frac{1}{2}\gamma^5 \bar{H_\alpha}^\tau\big)\big(\bar{F}_{\tau\beta}-\frac{1}{2}\gamma^5 \bar{H}_{\tau\beta}\big)\Big\rbrace \phi_{\nu}\\
			&\quad\qquad\qquad +\frac{1}{4}\bar{\psi}_{\mu}\Big\lbrace\gamma^\alpha\gamma^\rho\gamma^\mu\big(D_\rho \bar{F_\alpha}^\nu + \frac{1}{2}\gamma^5 D_\rho \bar{H_\alpha}^\nu \big)\\
			&\quad\qquad\qquad -\big(D_\rho \bar{F^\mu}_\alpha + \frac{1}{2}\gamma^5 D_\rho \bar{H^\mu}_\alpha\big)\gamma^\nu\gamma^\rho\gamma^\alpha \Big\rbrace \phi_{\nu}\\
			&\quad\qquad\qquad -\frac{1}{4}\bar{\phi}_{\mu}\Big\lbrace\gamma^\alpha\gamma^\rho\gamma^\mu\big(D_\rho \bar{F_\alpha}^\nu + \frac{1}{2}\gamma^5 D_\rho \bar{H_\alpha}^\nu \big)\\
			&\quad\qquad\qquad -\big(D_\rho \bar{F^\mu}_\alpha + \frac{1}{2}\gamma^5 D_\rho \bar{H^\mu}_\alpha\big)\gamma^\nu\gamma^\rho\gamma^\alpha \Big\rbrace \psi_{\nu}.
		\end{split}
	\end{align}
}
Then we have, 
{
	\allowdisplaybreaks
	\begin{subequations}
		\begin{align}
			{E^{\psi_\mu}}_{\psi_\nu}={E^{\phi_\mu}}_{\phi_\nu} &= \underbrace{- \mathbb{I}_4 {R^\mu}_\nu+\frac{1}{2}\gamma^\alpha\gamma^\beta {R^{\mu}}_{\nu\alpha\beta}-\frac{1}{2}\gamma^\mu\gamma^\alpha R_{\nu\alpha} +\frac{1}{2}\gamma_\nu\gamma^\alpha {R^\mu}_\alpha}_{X_1}\nonumber\\
			&\quad +\underbrace{\frac{1}{2}\gamma^\alpha\gamma^\beta\big(\bar{F}_{\alpha\beta}-\frac{1}{2}\gamma^5\bar{H}_{\alpha\beta}\big)\big(\bar{F^\mu}_\nu+\frac{1}{2}\gamma^5 \bar{H^\mu}_\nu\big)}_{X_2}\nonumber\\
			&\quad +\underbrace{\frac{1}{4}g^\mu_\nu\gamma^\alpha\gamma^\beta\big(\bar{F_\alpha}^\tau-\frac{1}{2}\gamma^5 \bar{H_\alpha}^\tau\big)\big(\bar{F}_{\tau\beta}-\frac{1}{2}\gamma^5 \bar{H}_{\tau\beta}\big)}_{X_3},\\
			{E^{\psi_\mu}}_{\phi_\nu}=-{E^{\phi_\mu}}_{\psi_\nu} &= \underbrace{\frac{1}{4}\gamma^\alpha\gamma^\rho\gamma^\mu\big(D_\rho \bar{F}_{\alpha\nu}+ \frac{1}{2}\gamma^5 D_\rho \bar{H_\alpha}_{\nu} \big)}_{X_4}\nonumber\\
			&\quad \underbrace{-\frac{1}{4}\big(D_\rho \bar{F^\mu}_\alpha + \frac{1}{2}\gamma^5 D_\rho \bar{H^\mu}_\alpha\big)\gamma_\nu\gamma^\rho\gamma^\alpha}_{X_5},\\
			{E^{\psi_\nu}}_{\psi_\mu}={E^{\phi_\nu}}_{\phi_\mu} &= \underbrace{- \mathbb{I}_4 {R^\nu}_\mu+\frac{1}{2}\gamma^\theta\gamma^\varphi {R^{\nu}}_{\mu\theta\varphi}-\frac{1}{2}\gamma^\nu\gamma^\theta R_{\mu\theta} +\frac{1}{2}\gamma_\mu\gamma^\theta {R^\nu}_\theta}_{Y_1}\nonumber\\
			&\quad +\underbrace{\frac{1}{2}\gamma^\theta\gamma^\varphi\big(\bar{F}_{\theta\varphi}-\frac{1}{2}\gamma^5\bar{H}_{\theta\varphi}\big)\big(\bar{F^\nu}_\mu+\frac{1}{2}\gamma^5 \bar{H^\nu}_\mu\big)}_{Y_2}\nonumber\\
			&\quad +\underbrace{\frac{1}{4}\bar{g}^\nu_\mu\gamma^\theta\gamma^\varphi\big(\bar{F_\theta}^\sigma-\frac{1}{2}\gamma^5 \bar{H_\theta}^\sigma\big)\big(\bar{F}_{\sigma\varphi}-\frac{1}{2}\gamma^5 \bar{H}_{\sigma\varphi}\big)}_{Y_3},\\
			{E^{\phi_\nu}}_{\psi_\mu}=-{E^{\psi_\nu}}_{\phi_\mu} &= \underbrace{-\frac{1}{4}\gamma^\beta\gamma^\sigma\gamma^\nu\big(D_\sigma \bar{F}_{\beta\mu}+ \frac{1}{2}\gamma^5 D_\sigma \bar{H_\beta}_{\mu} \big)}_{Y_4}\nonumber\\
			&\quad +\underbrace{\frac{1}{4}\big(D_\sigma \bar{F^\nu}_\beta + \frac{1}{2}\gamma^5 D_\sigma \bar{H^\nu}_\beta\big)\gamma_\mu\gamma^\sigma\gamma^\beta}_{Y_5}.
		\end{align}
	\end{subequations}
}
Then,
\begin{equation*}
	\text{tr}(X_1)=0,\thinspace \text{tr}(X_2)=0,\thinspace \text{tr}(X_3)=-8\bar{F}_{\mu\nu}\bar{F}^{\mu\nu},\thinspace \text{tr}(X_4)=0, \thinspace \text{tr}(X_5)=0.
\end{equation*}
So,
\begin{align}
	\begin{split}
		&{E^{\psi_\mu}}_{\psi_\mu}={E^{\phi_\mu}}_{\phi_\mu}=\sum_{i=1}^3 \text{tr}(X_i)= -8\bar{F}_{\mu\nu}\bar{F}^{\mu\nu},\\
		& {E^{\psi_\mu}}_{\phi_\mu}=-{E^{\phi_\mu}}_{\psi_\mu} = \sum_{i=4}^5 \text{tr}(X_i) = 0.
	\end{split}
\end{align}
Also,
{
	\allowdisplaybreaks
	\begin{equation*}
		\begin{split}
			X_1Y_1 &= 2R_{\mu\nu\rho\sigma}R^{\mu\nu\rho\sigma},\\
			X_2Y_2 &= 16R_{\mu\nu}R^{\mu\nu},\\
			X_3Y_3 &= 8(\bar{F}_{\mu\nu}\bar{F}^{\mu\nu})^2\\
			&\enspace-4R_{\mu\nu}R^{\mu\nu},
		\end{split}
		\hspace{0.1in}
		\begin{split}
			&X_4Y_4 = 0,\\
			&X_5Y_5 = 0,\\
			&X_1Y_2=X_2Y_1= 0,
		\end{split}
		\hspace{0.1in}
		\begin{split}
			X_1Y_3=X_3Y_1 &= 0,\\
			X_2Y_3=X_3Y_2&= 0,\\
			X_4Y_5=X_5Y_4&= 2R_{\mu\nu}R^{\mu\nu}\\
			&\enspace -2R_{\mu\nu\rho\sigma}\bar{F}^{\mu\nu}\bar{F}^{\rho\sigma}.
		\end{split}
	\end{equation*}
}
Therefore,
\begin{align}
	\begin{split}
		{E^{\psi_\mu}}_{\psi_\nu}{E^{\psi_\nu}}_{\psi_\mu}={E^{\phi_\mu}}_{\phi_\nu}{E^{\phi_\nu}}_{\phi_\mu}&= \sum_{i,j=1}^3 X_i Y_j \\
		&= 2R_{\mu\nu\rho\sigma}R^{\mu\nu\rho\sigma}+ 12R_{\mu\nu}R^{\mu\nu}+8(\bar{F}_{\mu\nu}\bar{F}^{\mu\nu})^2,\\
		{E^{\psi_\mu}}_{\phi_\nu}{E^{\phi_\nu}}_{\psi_\mu}={E^{\phi_\mu}}_{\psi_\nu}{E^{\psi_\nu}}_{\phi_\mu}&= \sum_{i,j=4}^5 X_i Y_j\\
		&= 4R_{\mu\nu}R^{\mu\nu}-4R_{\mu\nu\rho\sigma}\bar{F}^{\mu\nu}\bar{F}^{\rho\sigma}.
	\end{split}
\end{align}
\subsubsection*{Calculating $\text{tr}(\Omega_{\rho\sigma}\Omega^{\rho\sigma})$ }
\begin{align}
	\begin{split}
		\text{tr}(\Omega_{\rho\sigma}\Omega^{\rho\sigma}) &= {(\Omega_{\rho\sigma})^{\psi_\mu}}_{\psi_\nu}{(\Omega^{\rho\sigma})^{\psi_\nu}}_{\psi_\mu}+ {(\Omega_{\rho\sigma})^{\phi_\mu}}_{\phi_\nu}{(\Omega^{\rho\sigma})^{\phi_\nu}}_{\phi_\mu}\\
		&\quad + {(\Omega_{\rho\sigma})^{\psi_\mu}}_{\phi_\nu}{(\Omega^{\rho\sigma})^{\phi_\nu}}_{\psi_\mu}+ {(\Omega_{\rho\sigma})^{\phi_\mu}}_{\psi_\nu}{(\Omega^{\rho\sigma})^{\psi_\nu}}_{\phi_\mu},
	\end{split}
\end{align}
We have,
{
	\allowdisplaybreaks
	\begin{subequations}
		\begin{align}
			&{(\Omega_{\rho\sigma})^{\psi_\mu}}_{\psi_\nu}={(\Omega_{\rho\sigma})^{\phi_\mu}}_{\phi_\nu} = \mathbb{I}_4{R^{\mu}}_{\nu\rho\sigma}+\frac{1}{4}\bar{g}^\mu_\nu\gamma^\theta\gamma^\varphi R_{\rho\sigma\theta\varphi} + {[\omega_\rho,\omega_\sigma]^{\psi_{\mu}}}_{\psi_{\nu}},\\
			&{(\Omega_{\rho\sigma})^{\psi_\mu}}_{\phi_\nu}=-{(\Omega_{\rho\sigma})^{\phi_\mu}}_{\psi_\nu} ={D_{[\rho} {\omega_{\sigma]}}^{\psi_{\mu}}}_{\phi_{\nu}},\\
			&{(\Omega^{\rho\sigma})^{\psi_\nu}}_{\psi_\mu}={(\Omega^{\rho\sigma})^{\phi_\nu}}_{\phi_\mu} =\mathbb{I}_4{{R^{\nu}}_{\mu}}^{\rho\sigma}+\frac{1}{4}g^\nu_\mu\gamma^{\theta^\prime}\gamma^{\varphi^\prime} {R^{\rho\sigma}}_{\theta^\prime\varphi^\prime} + {[\omega^\rho,\omega^\sigma]^{\psi_{\nu}}}_{\psi_{\mu}},\\
			&{(\Omega^{\rho\sigma})^{\phi_\nu}}_{\psi_\mu}=-{(\Omega^{\rho\sigma})^{\psi_\nu}}_{\phi_\mu} = {D^{[\rho} {\omega^{\sigma]}}^{\phi_{\nu}}}_{\psi_{\mu}}.
		\end{align}
	\end{subequations}
}
So,
{
	\allowdisplaybreaks
	\begin{align}
		\begin{split}
			{(\Omega_{\rho\sigma})^{\psi_\mu}}_{\psi_\nu}{(\Omega^{\rho\sigma})^{\psi_\nu}}_{\psi_\mu} &= \big(\underbrace{\mathbb{I}_4{R^{\mu}}_{\nu\rho\sigma}+\frac{1}{4}\bar{g}^\mu_\nu\gamma^\theta\gamma^\varphi R_{\rho\sigma\theta\varphi}\big)\big(\mathbb{I}_4{{R^{\nu}}_{\mu}}^{\rho\sigma}+\frac{1}{4}\bar{g}^\nu_\mu\gamma^{\theta^\prime}\gamma^{\varphi^\prime} {R^{\rho\sigma}}_{\theta^\prime\varphi^\prime}}_{\textbf{Term1}} \big)\\
			&\qquad +\underbrace{2{R^{\mu}}_{\nu\rho\sigma}{[\omega^\rho,\omega^\sigma]^{\psi_{\nu}}}_{\psi_{\mu}}}_{\textbf{Term2}}+\underbrace{\frac{1}{2}\gamma^\theta\gamma^\varphi R_{\rho\sigma\theta\varphi}{[\omega^\rho,\omega^\sigma]^{\psi_{\mu}}}_{\psi_{\mu}}}_{\textbf{Term3}}\\
			&\qquad + \underbrace{{[\omega_\rho,\omega_\sigma]^{\psi_{\mu}}}_{\psi_{\nu}}{[\omega^\rho,\omega^\sigma]^{\psi_{\nu}}}_{\psi_{\mu}}}_{\textbf{Term4}}.
		\end{split}
	\end{align}
}
We compute,
\begin{enumerate}
	\item {
		\allowdisplaybreaks
		\begin{align}
			\textbf{Term1} &=\big(\mathbb{I}_4{R^{\mu}}_{\nu\rho\sigma}+\frac{1}{4}\bar{g}^\mu_\nu\gamma^\theta\gamma^\varphi R_{\rho\sigma\theta\varphi}\big)\big(\mathbb{I}_4{{R^{\nu}}_{\mu}}^{\rho\sigma}+\frac{1}{4}\bar{g}^\nu_\mu\gamma^{\theta^\prime}\gamma^{\varphi^\prime} {R^{\rho\sigma}}_{\theta^\prime\varphi^\prime}\big)\nonumber\\
			&= -6R_{\mu\nu\rho\sigma}R^{\mu\nu\rho\sigma},
	\end{align}}
	\item {
		\allowdisplaybreaks
		\begin{align*}
			\textbf{Term2} &= 2{R^{\mu}}_{\nu\rho\sigma}{[\omega^\rho,\omega^\sigma]^{\psi_{\nu}}}_{\psi_{\mu}}\nonumber\\
			& = 4{R^{\mu}}_{\nu\rho\sigma}{(\omega^\rho)^{\psi_\nu}}_{\phi_\tau}{(\omega^\sigma)^{\phi_\tau}}_{\psi_\mu}\\\nonumber
			&= 4\underbrace{{R^{\mu}}_{\nu\rho\sigma}}_{\text{X}}\Big(\underbrace{ \frac{1}{4}\gamma^\alpha\gamma^\rho\gamma^\nu({\bar{F}_{\alpha\tau}}+\frac{1}{2}\gamma^5{\bar{H}_{\alpha\tau}})}_{M_1} + \underbrace{\frac{1}{4}(\bar{F^\nu}_\alpha+\frac{1}{2}\gamma^5\bar{H^\nu}_\alpha)\gamma_\tau\gamma^\rho\gamma^\alpha}_{M_2}\Big)\\
			&\enspace \times \Big( \underbrace{-\frac{1}{4}\gamma^\beta\gamma^\sigma\gamma^\tau(\bar{F}_{\beta\mu}+\frac{1}{2}\gamma^5 \bar{H}_{\beta\mu})}_{N_1} \underbrace{- \frac{1}{4}(\bar{F^\tau}_\beta+\frac{1}{2}\gamma^5 \bar{H^\tau}_\beta)\gamma_\mu\gamma^\sigma\gamma^\beta}_{N_2}\Big).
	\end{align*}}
	Then, 
	\begin{equation*}
		XM_1N_1 =0,\thinspace XM_2N_2=0,\thinspace XM_1N_2=0,\thinspace XM_2N_1=0.
	\end{equation*}
	So,
	\begin{equation}
		\textbf{Term2} = 4\sum_{i,j=1}^2 XM_iN_j=0,
	\end{equation}
	\item {
		\allowdisplaybreaks
		\begin{align*}
			\textbf{Term3} &= \frac{1}{2}\gamma^\theta\gamma^\varphi R_{\rho\sigma\theta\varphi}{[\omega^\rho,\omega^\sigma]^{\psi_{\mu}}}_{\psi_{\mu}}\nonumber\\
			&= \gamma^\theta\gamma^\varphi R_{\rho\sigma\theta\varphi}{(\omega^\rho)^{\psi_\mu}}_{\phi_\tau}{(\omega^\sigma)^{\phi_\tau}}_{\psi_\mu}\nonumber\\
			&= \underbrace{\gamma^\theta\gamma^\varphi R_{\rho\sigma\theta\varphi}}_{Y}\Big( \underbrace{\frac{1}{4}\gamma^\alpha\gamma^\rho\gamma^\mu({\bar{F}_{\alpha\tau}}+\frac{1}{2}\gamma^5{\bar{H}_{\alpha\tau}})}_{U_1} + \underbrace{\frac{1}{4}(\bar{F^\mu}_\alpha+\frac{1}{2}\gamma^5 \bar{H^\mu}_\alpha)\gamma_\tau\gamma^\rho\gamma^\alpha}_{U_2}\Big)\\
			&\enspace \times\Big( \underbrace{-\frac{1}{4}\gamma^\beta\gamma^\sigma\gamma^\tau(\bar{F}_{\beta\mu}+\frac{1}{2}\gamma^5 \bar{H}_{\beta\mu})}_{V_1} \underbrace{- \frac{1}{4}(\bar{F^\tau}_\beta+\frac{1}{2}\gamma^5 \bar{H^\tau}_\beta)\gamma_\mu\gamma^\sigma\gamma^\beta}_{V_2}\Big).
	\end{align*}}
	Then, 
	\begin{equation*}
		YU_1V_1=0,\thinspace YU_2V_2=0, \thinspace YU_1V_2=0,\thinspace YU_2V_1= 0.
	\end{equation*}
	So,
	\begin{equation}
		\textbf{Term3}= \sum_{i,j=1}^2 YU_iV_j = 0,
	\end{equation}
	\item {
		\allowdisplaybreaks
		\begin{align*}
			\textbf{Term4} &= {[\omega_\rho,\omega_\sigma]^{\psi_{\mu}}}_{\psi_{\nu}}{[\omega^\rho,\omega^\sigma]^{\psi_{\nu}}}_{\psi_{\mu}}\nonumber\\
			&= 2\underbrace{{(\omega_\rho)^{\psi_\mu}}_{\phi_\tau}{(\omega_\sigma)^{\phi_\tau}}_{\psi_\nu}}_{T_1}\Big(\underbrace{{(\omega^\rho)^{\psi_\nu}}_{\phi_\lambda}{(\omega^\sigma)^{\phi_\lambda}}_{\psi_\mu}}_{T_2}-\underbrace{{(\omega^\sigma)^{\psi_\nu}}_{\phi_\lambda}{(\omega^\rho)^{\phi_\lambda}}_{\psi_\mu}}_{T_3} \Big),\\
		\end{align*}
	}
	where
	{
		\allowdisplaybreaks
		\begin{align*}
			T_1 &= \Big(\underbrace{ \frac{1}{4}\gamma^\alpha\gamma_\rho\gamma^\mu({\bar{F}_{\alpha\tau}}+\frac{1}{2}\gamma^5{\bar{H}_{\alpha\tau}})}_{X_1} + \underbrace{\frac{1}{4}(\bar{F^\mu}_\alpha+\frac{1}{2}\gamma^5\bar{H^\mu}_\alpha)\gamma_\tau\gamma_\rho\gamma^\alpha}_{X_2}\Big)\\
			&\enspace \times\Big( \underbrace{-\frac{1}{4}\gamma^\beta\gamma_\sigma\gamma^\tau(\bar{F}_{\beta\nu}+\frac{1}{2}\gamma^5 \bar{H}_{\beta\nu})}_{Y_1} \underbrace{- \frac{1}{4}(\bar{F^\tau}_\beta+\frac{1}{2}\gamma^5 \bar{H^\tau}_\beta)\gamma_\nu\gamma_\sigma\gamma^\beta}_{Y_2}\Big),\\
			T_2 &= \Big(\underbrace{ \frac{1}{4}\gamma^\theta\gamma^\rho\gamma^\nu({\bar{F}_{\theta\lambda}}+\frac{1}{2}\gamma^5{\bar{H}_{\theta\lambda}})}_{A_1} + \underbrace{\frac{1}{4}(\bar{F^\nu}_\theta+\frac{1}{2}\gamma^5 \bar{H^\nu}_\theta)\gamma_\lambda\gamma^\rho\gamma^\theta}_{A_2}\Big)\\
			&\enspace \times \Big( \underbrace{-\frac{1}{4}\gamma^\varphi\gamma^\sigma\gamma^\lambda(\bar{F}_{\varphi\mu}+\frac{1}{2}\gamma^5 \bar{H}_{\varphi\mu})}_{B_1} \underbrace{- \frac{1}{4}(\bar{F^\lambda}_\varphi+\frac{1}{2}\gamma^5 \bar{H^\lambda}_\varphi)\gamma_\mu\gamma^\sigma\gamma^\varphi}_{B_2}\Big),\\
			T_3 &= \Big(\underbrace{ \frac{1}{4}\gamma^\theta\gamma^\sigma\gamma^\nu({\bar{F}_{\theta\lambda}}+\frac{1}{2}\gamma^5{\bar{H}_{\theta\lambda}})}_{C_1} + \underbrace{\frac{1}{4}(\bar{F^\nu}_\theta+\frac{1}{2}\gamma^5 \bar{H^\nu}_\theta)\gamma_\lambda\gamma^\sigma\gamma^\theta}_{C_2}\Big)\\
			&\enspace \times \Big( \underbrace{-\frac{1}{4}\gamma^\varphi\gamma^\rho\gamma^\lambda(\bar{F}_{\varphi\mu}+\frac{1}{2}\gamma^5 \bar{H}_{\varphi\mu})}_{D_1} \underbrace{- \frac{1}{4}(\bar{F^\lambda}_\varphi+\frac{1}{2}\gamma^5 \bar{H^\lambda}_\varphi)\gamma_\mu\gamma^\rho\gamma^\varphi}_{D_2}\Big)
	\end{align*}}
	Then,
	{
		\allowdisplaybreaks
		\begin{equation*}
			\begin{split}
				&X_1Y_1A_1B_1 = -2R_{\mu\nu}R^{\mu\nu},\\
				&X_1Y_1A_2B_2= -2R_{\mu\nu}R^{\mu\nu},\\
				&X_1Y_1A_1B_2=0,\\
				&X_1Y_1A_2B_1=0,\\
				&X_2Y_2A_1B_1 =-2R_{\mu\nu}R^{\mu\nu},\\
				&X_2Y_2A_2B_2= -2R_{\mu\nu}R^{\mu\nu},\\
			\end{split}
			\hspace{0.1in}
			\begin{split}
				X_2Y_2A_1B_2&=0,\\
				X_2Y_2A_2B_1&=0,\\
				X_1Y_2A_1B_1 &=0,\\
				X_1Y_2A_2B_2&= 0,\\
				X_1Y_2A_1B_2&=2R_{\mu\nu}R^{\mu\nu}\\
				&\quad-4(\bar{F}_{\mu\nu}\bar{F}^{\mu\nu})^2,\\
			\end{split}
			\hspace{0.1in}
			\begin{split}
				X_1Y_2A_2B_1&=-2R_{\mu\nu}R^{\mu\nu},\\
				X_2Y_1A_1B_1 &= 0,\\
				X_2Y_1A_2B_2 &= 0,\\
				X_2Y_1A_1B_2 &= -2R_{\mu\nu}R^{\mu\nu},\\
				X_2Y_1A_2B_1 &=2R_{\mu\nu}R^{\mu\nu}\\
				&\quad-4(\bar{F}_{\mu\nu}\bar{F}^{\mu\nu})^2.
			\end{split}
	\end{equation*}}
	So,
	\begin{equation*}
		T_1T_2= \sum_{i,j,k,l=1}^2 X_iY_jA_kB_l = -8R_{\mu\nu}R^{\mu\nu}-8(\bar{F}_{\mu\nu}\bar{F}^{\mu\nu})^2.
	\end{equation*}
	Also,
	{
		\allowdisplaybreaks
		\begin{equation*}
			\begin{split}
				&X_1Y_1C_1D_1 = 4R_{\mu\nu}R^{\mu\nu},\\
				&X_1Y_1C_2D_2= 4R_{\mu\nu}R^{\mu\nu},\\
				&X_1Y_1C_1D_2=0,\\
				&X_1Y_1C_2D_1=0,\\
				&X_2Y_2C_1D_1 =4R_{\mu\nu}R^{\mu\nu},\\
				&X_2Y_2C_2D_2= 4R_{\mu\nu}R^{\mu\nu},\\
			\end{split}
			\hspace{0.1in}
			\begin{split}
				X_2Y_2C_1D_2&=0,\\
				X_2Y_2C_2D_1&=0,\\
				X_1Y_2C_1D_1 &=0,\\
				X_1Y_2C_2D_2&= 0,\\
				X_1Y_2C_1D_2&=-4R_{\mu\nu}R^{\mu\nu}\\
				&\quad+8(\bar{F}_{\mu\nu}\bar{F}^{\mu\nu})^2,\\
			\end{split}
			\hspace{0.1in}
			\begin{split}
				X_1Y_2C_2D_1&=4R_{\mu\nu}R^{\mu\nu},\\
				X_2Y_1C_1D_1 &= 0,\\
				X_2Y_1C_2D_2 &= 0,\\
				X_2Y_1C_1D_2 &= 4R_{\mu\nu}R^{\mu\nu},\\
				X_2Y_1C_2D_1 &=-4R_{\mu\nu}R^{\mu\nu}\\
				&\quad+8(\bar{F}_{\mu\nu}\bar{F}^{\mu\nu})^2.
			\end{split}
	\end{equation*}}
	So,
	\begin{equation*}
		T_1T_3= \sum_{i,j,k,l=1}^2 X_iY_jC_kD_l = 16R_{\mu\nu}R^{\mu\nu}+16(\bar{F}_{\mu\nu}\bar{F}^{\mu\nu})^2.
	\end{equation*}
	Hence,
	\begin{equation}
		\textbf{Term4}= 2(T_1T_2-T_1T_3)= -48R_{\mu\nu}R^{\mu\nu}-48(\bar{F}_{\mu\nu}\bar{F}^{\mu\nu})^2.
	\end{equation}
\end{enumerate}
Finally, we have
\begin{align}
	{(\Omega_{\rho\sigma})^{\psi_\mu}}_{\psi_\nu}{(\Omega^{\rho\sigma})^{\psi_\nu}}_{\psi_\mu}&={(\Omega_{\rho\sigma})^{\phi_\mu}}_{\phi_\nu}{(\Omega^{\rho\sigma})^{\phi_\nu}}_{\phi_\mu}\nonumber\\
	&= \textbf{Term1}+\textbf{Term2}+\textbf{Term3}+\textbf{Term4}\nonumber\\
	&= -6R_{\mu\nu\rho\sigma}R^{\mu\nu\rho\sigma}-48R_{\mu\nu}R^{\mu\nu}-48(\bar{F}_{\mu\nu}\bar{F}^{\mu\nu})^2.
\end{align}
Again,
\begin{align}
	{(\Omega_{\rho\sigma})^{\psi_\mu}}_{\phi_\nu}{(\Omega^{\rho\sigma})^{\phi_\nu}}_{\psi_\mu}&= {D_{[\rho} {\omega_{\sigma]}}^{\psi_{\mu}}}_{\phi_{\nu}}{D^{[\rho} {\omega^{\sigma]}}^{\phi_{\nu}}}_{\psi_{\mu}}\nonumber\\
	&= 2\underbrace{{{(D_\rho\omega_\sigma)}^{\psi_\mu}}_{\phi_\nu}}_{T_4}\Big(\underbrace{{{(D^\rho\omega^\sigma)}^{\phi_\nu}}_{\psi_\mu}}_{T_5}-\underbrace{{{(D^\sigma\omega^\rho)}^{\phi_\nu}}_{\psi_\mu}}_{T_6} \Big),
\end{align}
{
	\allowdisplaybreaks
	\begin{align*}
		T_4 &= \underbrace{ \frac{1}{4}\gamma^\alpha\gamma_\sigma\gamma^\mu(D_\rho{\bar{F}_{\alpha\nu}}+\frac{1}{2}\gamma^5 D_\rho{\bar{H}_{\alpha\nu}})}_{L_1} + \underbrace{\frac{1}{4}(D_\rho\bar{F^\mu}_\alpha+\frac{1}{2}\gamma^5 D_\rho\bar{H^\mu}_\alpha)\gamma_\nu\gamma_\sigma\gamma^\alpha}_{L_2},\\
		T_5 &=  \underbrace{-\frac{1}{4}\gamma^\beta\gamma^\sigma\gamma^\nu(D^\rho \bar{F}_{\beta\mu}+\frac{1}{2}\gamma^5 D^\rho \bar{H}_{\beta\mu})}_{P_1} \underbrace{- \frac{1}{4}(D^\rho\bar{F^\nu}_\beta+\frac{1}{2}\gamma^5 D^\rho\bar{H^\nu}_\beta)\gamma_\mu\gamma^\sigma\gamma^\beta}_{P_2},\\
		T_6 &=  \underbrace{-\frac{1}{4}\gamma^\beta\gamma^\rho\gamma^\nu(D^\sigma \bar{F}_{\beta\mu}+\frac{1}{2}\gamma^5 D^\sigma \bar{H}_{\beta\mu})}_{Q_1} \underbrace{- \frac{1}{4}(D^\sigma\bar{F^\nu}_\beta+\frac{1}{2}\gamma^5 D^\sigma\bar{H^\nu}_\beta)\gamma_\mu\gamma^\rho\gamma^\beta}_{Q_2}.
\end{align*}}
Then,
{
	\allowdisplaybreaks
	\begin{equation*}
		\begin{split}
			&L_1P_1= 0,\\
			&L_2P_2 =0,\\
			&L_1P_2 = 8R_{\mu\nu\rho\sigma}\bar{F}^{\mu\nu}\bar{F}^{\rho\sigma}-8R_{\mu\nu}R^{\mu\nu},\\
			&L_2P_1 = 8R_{\mu\nu\rho\sigma}\bar{F}^{\mu\nu}\bar{F}^{\rho\sigma}-8R_{\mu\nu}R^{\mu\nu},
		\end{split}
		\hspace{0.3in}
		\begin{split}
			&L_1Q_1= 0,\\
			&L_2Q_2 =0,\\
			&L_1Q_2 = 2R_{\mu\nu\rho\sigma}\bar{F}^{\mu\nu}\bar{F}^{\rho\sigma}-2R_{\mu\nu}R^{\mu\nu},\\
			&L_2Q_1 = 2R_{\mu\nu\rho\sigma}\bar{F}^{\mu\nu}\bar{F}^{\rho\sigma}-2R_{\mu\nu}R^{\mu\nu}.
		\end{split}
\end{equation*}}
So, 
\begin{align*}
	T_4T_5 &= \sum_{i,j=1}^2 L_iP_j = 16R_{\mu\nu\rho\sigma}\bar{F}^{\mu\nu}\bar{F}^{\rho\sigma}-16R_{\mu\nu}R^{\mu\nu},\\
	T_4T_6 &= \sum_{i,j=1}^2 L_iQ_j = 4R_{\mu\nu\rho\sigma}\bar{F}^{\mu\nu}\bar{F}^{\rho\sigma}-4R_{\mu\nu}R^{\mu\nu}.
\end{align*}
Hence, we have
\begin{align}
	{(\Omega_{\rho\sigma})^{\psi_\mu}}_{\phi_\nu}{(\Omega^{\rho\sigma})^{\phi_\nu}}_{\psi_\mu}={(\Omega_{\rho\sigma})^{\phi_\mu}}_{\psi_\nu}{(\Omega^{\rho\sigma})^{\psi_\nu}}_{\phi_\mu}&=2(T_4T_5-T_4T_6)\nonumber\\
	&= 24R_{\mu\nu\rho\sigma}\bar{F}^{\mu\nu}\bar{F}^{\rho\sigma}-24R_{\mu\nu}R^{\mu\nu}.
\end{align}


\end{document}